\documentclass[sn-mathphys]{sn-jnl}% Math and Physical Sciences Reference Style
\usepackage{bm}

\jyear{2021}%

\theoremstyle{thmstyleone}%
\theoremstyle{thmstyletwo}%

\theoremstyle{thmstylethree}%

\graphicspath{{fig-sec2/}{fig-sec3/}{fig-sec4/}{fig-sec5/}}
\begin{document}

\title[Production of polarized particle beams via ultraintense laser pulses]{Production of polarized particle beams via ultraintense laser pulses}

\author[1]{\fnm{Ting} \sur{Sun}}\email{small\_black@stu.xjtu.edu.cn}
\author*[1]{\fnm{Qian} \sur{Zhao}}\email{zhaoq2019@xjtu.edu.cn}
\author[1]{\fnm{Kun} \sur{Xue}}\email{xkjy56@stu.xjtu.edu.cn}
\author[1]{\fnm{Zhi-Wei} \sur{Lu}}\email{luzhw2020@stu.xjtu.edu.cn}
\author[2]{\fnm{Liang-Liang} \sur{Ji}}\email{jill@siom.ac.cn}
\author[1]{\fnm{Feng} \sur{Wan}}\email{wanfeng@xjtu.edu.cn}
\author[1]{\fnm{Yu} \sur{Wang}}\email{wy1995@stu.xjtu.edu.cn}
\author[3]{\fnm{Yousef I.} \sur{Salamin}}\email{ysalamin@aus.edu}
\author*[1]{\fnm{Jian-Xing} \sur{Li}}\email{jianxing@xjtu.edu.cn}

\affil[1]{\orgdiv{Ministry of Education Key Laboratory for Nonequilibrium Synthesis and Modulation of Condensed Matter, School of Physics}, \orgname{Xi'an Jiaotong University}, \orgaddress{\city{Xi'an}, \postcode{710049}, \state{Shaanxi}, \country{China}}}
\affil[2]{\orgdiv{State Key Laboratory of High Field Laser Physics}, \orgname{Shanghai Institute of Optics and Fine Mechanics, Chinese Academy of Sciences}, \orgaddress{\city{Shanghai}, \postcode{201800}, \country{China}}}
\affil[3]{\orgdiv{Department of Physics}, \orgname{American University of Sharjah}, \orgaddress{\city{Sharjah}, \postcode{POB 26666},  \country{United Arab Emirates}}}


\abstract{High-energy spin-polarized electron, positron, and $\gamma$-photon beams have many significant applications in the study of material properties, nuclear structure, particle physics, and high-energy astrophysics. Thus, efficient production of such polarized beams attracts a broad spectrum of research interests.	This is driven mainly by the rapid advancements in ultrashort and ultraintense laser technology. Currently available	laser pulses can achieve peak intensities in the range of $10^{22}-10^{23}$ Wcm$^{-2}$, with pulse durations of tens of femtoseconds. The dynamics of particles in laser fields of the available intensities is dominated by quantum electrodynamics (QED) and the interaction mechanisms have reached regimes spanned by nonlinear multiphoton absorbtion (strong-field QED processes).	 In strong-field QED processes, the scattering cross sections obviously depend on the spin and polarization of the particles, and the spin-dependent photon emission and the radiation-reaction effects can be utilized to produce the polarized particles. An ultraintense laser-driven polarized particle source possesses the advantages of high-brilliance and compactness, which could open the way for the unexplored aspects in a range of researches. In this work, we briefly review the seminal conclusions from the  study of the polarization effects in strong-field QED processes, as well as the progress made by recent proposals for production of the polarized particles by laser-beam or laser-plasma interactions.
}




\maketitle

\section{Introduction}\label{introduction}
Spin-polarized particles are employed extensively to investigate the properties of materials, in atomic and molecular structure investigations \cite{gay2009physics,schultz1988interaction,danielson2015plasma}, in nuclear structure studies \cite{abe1995precision,uggerhoj2005the,alexakhin2007deuteron}, and in high-energy physics \cite{moortgat2008polarized}. Being chiral, spin-polarized relativistic particles are ideally suited for selective enhancement or suppression of specific reaction channels which, in turn, makes  them better at characterizing the quantum numbers and chiral couplings of the new particles \cite{barish2013international}. For instance, polarized $\gamma$ photons with tens of MeV energy can be used to excite polarization-dependent photofission of the nucleus in the giant dipole resonance \cite{speth1981giant}, while polarized $\gamma$ photons with GeV energy play crucial roles in meson photoproduction \cite{akbar2017measurement} and the test of vacuum birefringence \cite{bragin2017high}. In astrophysics polarization of the $\gamma$ photons provides detailed insight into the $\gamma$-photon emission mechanism and the properties of dark matter \cite{laurent2011polarized,boehm2017circular}. Furthermore, spin-polarized lepton beams are used in investigations of the spin dependence of the fundamental interactions and the violation of symmetries such as parity \cite{schlimme2013measurement}. Additionally, they play a central role in precise measurements pertaining to the spin-dependent processes that focus attention on physics beyond the standard model. Such measurements can compete with direct searches at the high-energy accelerators \cite{jefferson2018precision}.

Conventional methods of producing high-energy polarized $\gamma$ photons include linear Compton scattering \cite{howell2021international} and bremsstrahlung \cite{olsen1959photon,kuraev2010bremsstrahlung}. The former employs unpolarized electron beams in which the emitted $\gamma$-photon polarization is determined by the driving laser polarization, because the radiation formation length is much longer than the laser wavelength \cite{Baier1973,ritus1985quantum,sokolov1986radiation,khokonov2010length}. Because of the relatively small scattering cross-section of the linear Compton scattering (around $10^{-3}$ barns$/$MeV), luminosity of the electron-photon collision is rather low \cite{omori2006efficient}. The collision luminosity can be increased by using high-intensity lasers, but in this case the interaction moves into the nonlinear regime. Many theoretical and experimental investigations have recently been conducted into how to acquire brilliant high-energy $\gamma$-rays, even with spin or orbital angular momentum in the nonlinear regime.

In incoherent bremsstrahlung, an electron beam is passed through a thin metal target and radiates in the presence of the Coulomb field near the nucleus \cite{giulietti2008intense,albert2016applications}. Here, the circularly-polarized (CP) $\gamma$-photons are generated by longitudinally spin-polarized (LSP) electrons \cite{ugger2005the,timm1969coherent} while the linearly-polarized (LP) $\gamma$-photons cannot  be generated  because of the large scattering angle \cite{baier1998electromagnetic}. LP $\gamma$-photons can be obtained by coherent bremsstrahlung when an electron beam is passed through a crystal, which gets periodically disturbed by the nucleus. The damage threshold of the crystal, however, places limits on the energy of the electron beam and the flux of the $\gamma$-rays.

The usual sources of polarized leptons are obtained in a storage ring via the Sokolov-Ternov (ST) effect \cite{sokolov1986radiation} and Bethe-Heitler  pair production. The transversely spin-polarized (TSP) lepton beams directly obtained via the ST effect require long polarization times, since the corresponding static magnetic fields are relatively weak. Longitudinally spin-polarized (LSP) leptons can be produced in a Bethe-Heitler process via high-energy circularly polarized $\gamma$-photons interacting with a high-Z target \cite{abbott2016production}. In such a process, the low luminosity of the $\gamma$-photon beam is compensated by meeting the high repetition rate requirement to yield a dense positron beam for applications \cite{variola2014advanced}. Generally speaking, the transverse and longitudinal polarizations can be transformed into each other by a spin-rotator. This, however, has the drawbacks of demanding quasi-monoenergetic beams and being accompanied by a beam intensity reduction. To enable the spin-dependent sciences at the ever-demanding energy and brilliance, development of advanced polarized particle beam is considered essential.

The rapid development of ultraintense  ultrashort laser techniques have followed the advent of petawatt laser systems. Current state-of-the-art laser pulses can achieve peak intensities of about $10^{23}$ W/cm$^2$, with pulse durations of tens of femtoseconds and energy fluctuations of about 1\% \cite{danson2019petawatt,yoon2019achieving,yoon2021realization}. Polarized particles can be generated in laser-plasma acceleration experiments by focusing laser light on a  prepolarized target, or from the polarization build-up in strong-field QED interactions between the laser and particles.
Laser-plasma accelerators sustain ultrahigh acceleration gradients (exceeding 0.1 TeV/m) and can generate dense (tens-of-MeV) proton \cite{higginson2018near, mcilvenny2021selective} and (multi-GeV) electron beams \cite{gonsalves2019petawatt} in experiments utilizing high-density prepolarized gas targets \cite{sofikitis2018ultrahigh}. Theoretical investigations indicate that the prepolarized particles suffer acceptable levels of depolarization during
 acceleration in laser-driven plasma fields and yield dense polarized electron  \cite{wen2019polarized,wu2019polarizedNJP,wu2019polarizedPRE,nie2021situ} and  proton beams  \cite{gong2020energetic,jin2020spin,li2021polarized}.

 TSP electron and positron beams can  be directly produced by strong-field QED processes, such as nonlinear Compton scattering (NLC)  [nonlinear Breit-Wheeler (NBW)  pair production] in elliptically polarized \cite{li2019ultrarelativistic,wan2020ultrarelativistic} or bichromatic laser pulses \cite{seipt2019ultrafast,song2019spin,chen2019polarized,liu2020trapping}, due to the quantum radiative spin effects.  LSP beams can  be produced   via the helicity transfer from CP $\gamma$ photons in NBW  process  \cite{li2020production,xue2021generation},  which  are generally  pre-produced via Compton scattering \cite{phuoc2012all,li2020polarized} or bremsstrahlung  \cite{abbott2016production}.

The exact treatment of strong-field QED in the Furry picture is limited to solving the Dirac equation in an external field \cite{berestetskii1982quantum,salamin1993dirac,piazza2012extremely,fedotov2022advances}. Exact solutions are known only for some specific fields, such as the Coulomb field, the constant and homogeneous magnetic field, the plane electromagnetic wave field, and the constant crossed electric and magnetic fields. In the Furry picture, Volkov states are employed in investigation of the interaction of two {\it dressed} fermion lines in both the NLC and NBW (the first-order processes in quantum electrodynamical interactions of particles in a strong plane-wave background field). In the theory advanced by Nikishov and Ritus \cite{nikishov1964quantum}, the Volkov state wavefunction is normalized to {\it one particle per  unit volume},  which leads to a quasi-momentum for the electron, thus the calculations of NLC and NBW are handled by similar linear processes. The polarization-resolved NLC and NBW processes are obtained in this theory \cite{ritus1972radiative,bol2000spin,ivanov2004complete,ivanov2005complete,karlovets2011radiative,blackburn2020radiation}. Employing Volkov states, spin- and polarization-dependent probabilities of the NLC and NBW are calculated in the locally-constant-field-approximation (LCFA) regime \cite{king2013photon,seipt2018theory,seipt2020spin}, as well as in the fully strong-field QED theory \cite{king2020nonlinear,tang2020highly}. Lately, the probabilities of higher-order-in-$\alpha$ ($\alpha$ is fine-structure constant) strong-field QED processes are constructed in terms of Stokes vectors and Mueller matrices, applicable for general pulsed fields and general spin and polarization \cite{dinu2020approximation, torgrimsson2021loops}.

There is also the quasiclassical operator (QO) approach, developed by Bair and Katkov to study the strong-field QED processes under a variety of strong background fields \cite{berestetskii1982quantum,baier1998electromagnetic,baier2009recent}. The basic idea here is that two types of quantum effects are involved in the radiation emitted by the high-energy particle in an external field. The first effect is associated with the quantization of the motion of the particle in the field. The magnitude of noncommutativity of the particle's dynamical variables is of the order of $\hbar/(\varepsilon\rho)$, where $\varepsilon$ is the particle energy and $\rho$ is some length characterizing the motion of the charged particle in the external field. The second type of quantum effects is associated with the recoil of the particle following the emission of radiation and is of order $\hbar\omega/\varepsilon$, with $\omega$ the frequency of the radiated photon. Energy of the emitted photon depends on the value of the quantum parameter $\chi$. That energy is $\hbar\omega\sim\varepsilon$, when $\chi\gtrsim1$. Thus, the quantum effects of the first type are negligible for ultra-relativistic particle energies ($\gamma\gg1$). In this regime, the dynamics of particles can be considered to be quasiclassical. The polarized NLC and NBW processes are extensively investigated employing the QO approach in the LCFA regime \cite{li2020polarized,li2020production,xue2021generation,chen2022electron} and in the weakly nonlinear regime \cite{wistisen2019numerical,wistisen2020numerical}. It has the advantage of allowing for numerical solutions in arbitrary electromagnetic fields, which is favored in the investigation of interactions between ultraintense tightly-focused laser pulses and ultrarelativistic electron beams. A series of Monte Carlo (MC) algorithms based on the QO theory have been developed to study the spin and polarization effects in strong-field QED processes \cite{chen2019polarized,wan2020ultrarelativistic,li2020production,xue2021generation}. Some of those algorithms are also coupled to particle-in-cell (PIC) codes to study the QED of plasmas \cite{song2021spin,song2021dense}.

In this paper, work is reviewed that has been devoted to the acceleration of polarized particles and polarized strong-field first-order QED processes, namely, the NLC and NBW scattering driven by the strong laser field. The paper is organized as follows. The acceleration of pre-polarized electron and proton beams in a plasma-based accelerator is reviewed in Sect. \ref{sec-polacc}. The polarization effects and spin-polarization build-up of electrons in weakly- and strongly-NLC processes are summarized in Sect. \ref{sec-polele}. The polarization (angular momentum) build-up of $\gamma$-photons from laser polarization and electron spin in weakly- and strongly-NLC processes is briefly reviewed in Sect. \ref{sec-polgamm}. The spin-polarization build-up of positrons in the NBW process and proposals for producing polarized positrons are discussed in Sect. \ref{sec-polposi}. Finally, a summary of this work is given in Sect. \ref{sec:summ}.

\section{Acceleration of polarized particle beams in plasma-based accelerators}\label{sec-polacc}

Traditional methods of producing polarized electron beams via radiative polarization in storage rings or photoemmison from a gallium
arsenide (GaAs) cathode \cite{pierce1976photoemission,sokolov1986radiation} produce low currents and are limited to applications in atomic, nuclear, and particle physics. Laser- or beam-driven plasma wakefield acceleration sustains acceleration gradient in excess of 100 GV m$^{-1}$, in the blowout regime, and generates electron beams with kilo-ampere currents and ultrahigh brightness \cite{esarey2009physics,tajima2020wakefield}. Thus it is advantageous to produce the high-current polarized electron beams in a plasma-based accelerator. The polarized gas target can be produced by means of  ultraviolet (UV) photodissociation of a hydrogen halide \cite{rakitzis2003spin,rakitzis2004pulsed,sofikitis2008nanosecond,sofikitis2017highly,sofikitis2018ultrahigh}. Polarized hydrogen atoms with gas densities reaching $10^{19}$ cm$^{-3}$ have been obtained experimentally \cite{sofikitis2018ultrahigh}, which are suitable for wakefield acceleration of electron and proton beams. During the injection and acceleration of  polarized plasma electrons into the plasma wakefield, it is critical  to evaluate the depolarization of the electron beam, since the polarization is fully determined by the temporal evolution of the direction of the
spin ($\bm{s}$) of each electron in the beam.

\begin{figure}[!t]
	\setlength{\abovecaptionskip}{0.2cm}  	
	\centering\includegraphics[width=0.8\linewidth]{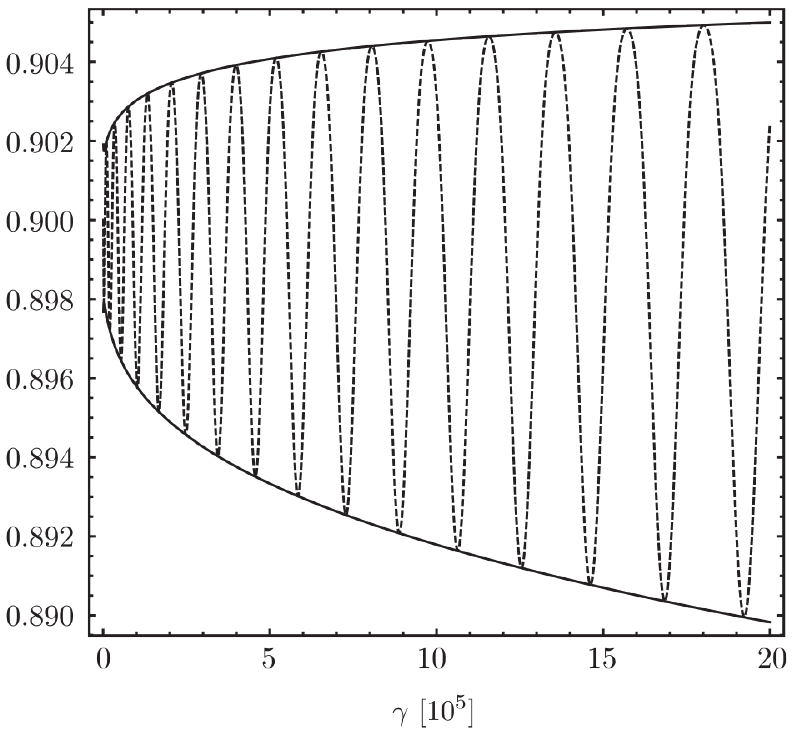}	
	\caption{Evolution of the longitudinal spin component $s_z$ as a function of $\gamma$ for a single electron with zero initial velocity,  calculated from Eq. (\ref{sz-evolve}) with $r_0=0.5c/\omega_p, s_{z0}=0.9,$ and $s_{r0}=0.1$. The solid and dashed lines denote the envelope of the oscillations and $s_z(t)$, respectively. \cite{vieira2011polarized}.}
	\label{fig2-1}
\end{figure}

In electric ($\bm{E}$) and magnetic ($\bm{B}$) fields, the direction of the electron spin is determined by the ST effect (spin flip) and spin precession according to the Thomas-Bargmann-Michel-Telegdi (T-BMT) equation. The ST effect gradually transits  the direction of electron spin to the antiparallel direction of static magnetic field after radiating photons, however, the typical polarization time ($\sim$$\mu$s scale) for electrons via ST effect is much larger than the acceleration duration ($\sim$ns scale), and the ST effect can be neglected in plasma wakefield \cite{wu2019polarizedNJP,thomas2020scaling}. Although the T-BMT equation is strictly valid when electric and magnetic fields are homogenous, it can still be used in the wakefield because the Stern-Gerlach forces can be neglected compared to the Lorentz forces and thus does not change the electron trajectories \cite{vieira2011polarized,wu2019polarizedNJP,thomas2020scaling}. Therefore, the spin dynamics of electrons in the wakefield are dominated by the T-BMT equation $d\bm{s}/dt=(\bm{\Omega}_T+\bm{\Omega}_a)\times\bm{s}$ where
\begin{eqnarray}\label{BMT}
\bm{\Omega}_T &=&\frac{e}{m_ec}\left(\frac{1}{\gamma}\bm{B}-\frac{1}{\gamma+1}\frac{\bm{\beta}}{c}\times\bm{E}\right),\\
\bm{\Omega}_a &=&a_e\frac{e}{m_e}\left(\bm{B}-\frac{\gamma}{\gamma+1}\bm{\beta}(\bm{\beta}\cdot\bm{B})-\frac{\bm{\beta}}{c}\times\bm{E}\right),
\end{eqnarray}
with $a_e\approx1.16\times10^{-3}$ the anomalous magnetic moment of electron, and $\gamma=1/\sqrt{1-\beta^2}$ the relativistic factor of electron. Considering the motion of an electron in the wakefield propagaring along $\hat{\bm{z}}$ direction, its spin vector with cylindrical coordinates $\bm{s}=(\bm{s}_r,\bm{s}_\phi,\bm{s}_z)$ is thus subjected to the radial component of wakefield $W_\perp=\bm{E}_r-c\bm{B}_\phi$ and evolves by T-BMT equation. Assuming the linearly radial and accelerating field (it is the situation for blowout regime of wake wave) in Eq. (\ref{BMT}), the analytical solution is obtained \cite{vieira2011polarized}
\begin{equation}\label{sz-evolve}
s_z[\gamma(t)]=\sqrt{1-s_{\phi0}^2}\sin{\left[r_0\Phi[\gamma(t)]+\arctan{\left(\frac{s_{z0}}{s_{r0}}\right)}\right]}
\end{equation}
where $s_{z0}$, $s_{r0}$ and $s_{\phi0}$ are initial spin components in cylindrical coordinate system, $\Phi[\gamma(t)]$ is the energy-dependent term for evolutional electron energy $\gamma(t)$. It shows that $s_z$ oscillates with the betatron frequency, and that the amplitude of the
oscillations depends directly on the electron energy due to $\Phi[\gamma(t)]$ (see Fig. \ref{fig2-1}). Thus a plasma accelerator can be efficiently used to accelerate polarized electron beams while preserving the high polarization.

Assuming the pre-polarized gas jet produced by photodissociation of a circularly polarized UV laser pulse, a number of proposals have been put forward for producing  high-current polarized electron beams via plasma wakefield acceleration. Wen \emph{et al}. employ fully three-dimensional (3D) PIC simulations to the laser wakefield acceleration (LWFA) with a pre-polarized plasma of a narrow density bump, which leads to the shock-front injection \cite{wen2019polarized}. In the 3D simulations, spins of the plasma electrons are initially directed along the $x$ axis (wakefield axis). The precession frequency $\bm{\Omega}_T(t)\propto\bm{B}_\phi(r)/\gamma$, where $\bm{B}_\phi$ is the azimuthal
magnetic field of the wake wave, determines evolution of the electron spin direction, which decreases steadily from its maximum as the sheath electron locates at the focusing region of the wakefield (where an electron trajectory has a relatively large off-axis radius $r$, implying a large $\bm{B}_\phi(r)$ and $\gamma\approx1$) to nearly zero as the sheath electron injects into the ion channel of the wake wave. Therefore, the symmetric precession implies a symmetric
spin spreading with $\langle s_y\rangle\approx\langle s_z\rangle\approx0$ and a rapid decrease of $\langle s_x\rangle$ at the injection stage. The final energetic electron beams carrying polarization and currents in the weakly nonlinear wakefield regime are $(90.6\%, 73.9\%, 53.5\%)$ and $(0.31,0.59,0.90)$ kA, corresponding to the respective values of the driving laser dimensionless intensity parameter $a_0=(1,1.1,1.2)$. Therefore, in order to avoid severe spin depolarization at the injection stage, the wakefield should be moderately strong, which limits both the accelerating gradient and charge of the injected electrons. Wu \emph{et al}. propose that a Laguerre-Gaussian laser-driven donut-shaped wakefield, which accelerates the annular electron beam, should be capable of suppressing the beam depolarization without sacrificing the injected electron charge \cite{wu2019polarizedNJP}. This can happen because the donut-shaped wakefield significantly lowers the electron areal density and the current density for a certain amount of beam charge, and thus reduces $B_\phi$ as the spin procession is strongly related to it. Depolarization of the injected electron beam could also be mitigated by the acceleration in the particle beam-driven wakefield at the beam-plasma interaction stage, because a typical strength of the beam field is about two orders of magnitude smaller than the laser field in LWFA \cite{wu2019polarizedPRE}. Furthermore, Nie \emph{et al}. employ a combination of time-dependent Schrödinger equation and 3D PIC simulations, in which the spin-polarized electrons can be produced \emph{in situ} while the Xe gas is ionized by a UV laser behind a beam driver. The polarized electrons are produced by ionization, injected inside a beam-driven plasma bubble with a strongly nonlinear wakefield, and rapidly accelerated to multi-GeV energies. The ionization injection guarantees that the injected polarized electrons have small orbit radii near the wake axis, where the transverse magnetic and electric fields of the wake are near zero, which minimizes both the beam emittance and depolarization due to spin precession.

Currently-available petawatt laser pulses make it possible to accelerate ion beams in a gas density plasma \cite{shen2007bubble,shen2009high,zhang2014proton,wan2019two,engin2019laser}, where the ion beam can be trapped and accelerated by a plasma bubble-channel formed by ponderomotive expulsion of the electrons, as a relativistic self-focusing effect in high-power laser-plasma interactions, when the laser power exceeds the critical value of $P_c=17n_c/n_0$ GW. Compared to conventional accelerators, the intensity of laser-driven proton beams can be increased by several orders of magnitude. Assuming the pre-polarized hydrogen chloride gas with density $10^{19}$ cm$^{-3}$, several theoretical investigations have recently demonstrated that spin-polarized proton beams can be generated in the laser interaction with a gas-jet while preserving high polarization ($>80\%$) \cite{hutzen2019polarized,buscher2019polarized,hutzen2020simulation,jin2020spin,gong2020energetic,li2021polarized,reichwein2021robustness}. 3D-PIC simulations have demonstrated that the pre-polarized protons undergo two-stage acceleration in a plasma bubble-channel. In the first stage, the protons are accelerated by longitudinal and radial space-charge fields. In the second stage, the focused protons in the filament are accelerated strongly by the strong longitudinal field generated through the transverse expansion of a magnetic field, according to Faraday’s law (magnetic vortex acceleration) \cite{jin2020spin}. When a petawatt laser pulse propagates through a compound plasma target, consisting of a double layer slab of near-critical-density carbon nanotube foams mixed with pre-polarized HCl gas,  a polarized proton beam can be accelerated to a maximum energy of 0.5 GeV with polarization as high as 94\% \cite{gong2020energetic}. Acceleration of spin-polarized protons is also considered in the acceleration regime consisting of a near-critical-density plasma target with pre-polarized proton and tritium ions, where the pre-polarized protons are initially accelerated by laser radiation pressure before injection and further acceleration in a bubble-like wakefield \cite{li2021polarized}.

\section{Polarization effects of electrons in nonlinear Compton scattering}\label{sec-polele}

\begin{figure}[!t]
	\setlength{\abovecaptionskip}{0.2cm}  	
	\centering\includegraphics[width=0.9\linewidth]{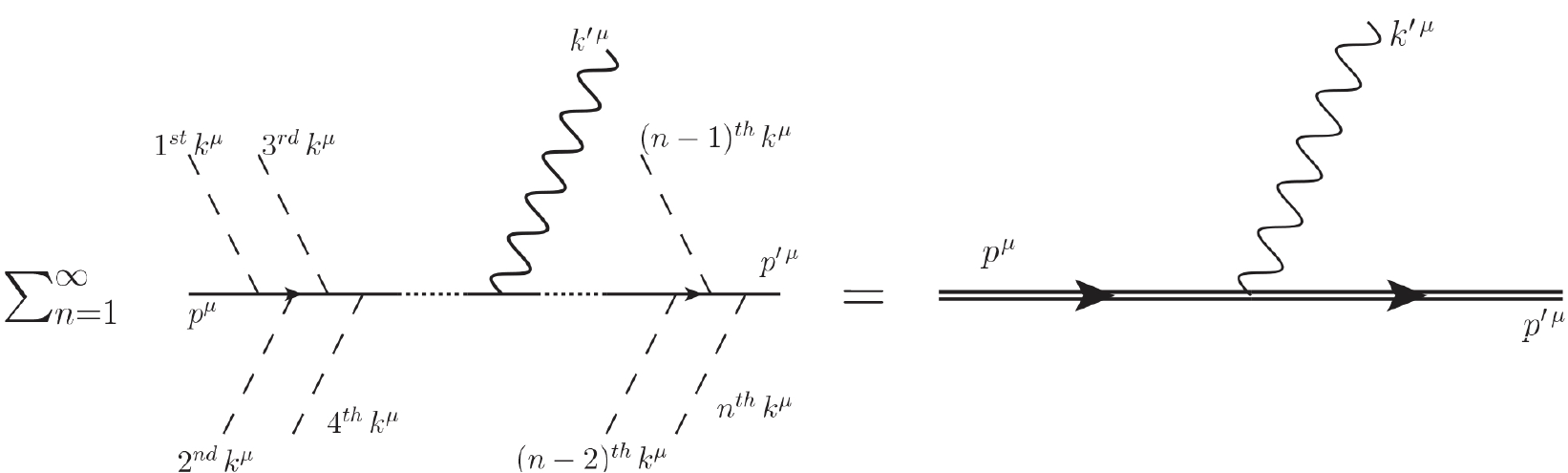}	
	\caption{Feynman diagrams of multiphoton Compton scattering drawn in a conventional QED picture (left) and in the Furry picture (right) \cite{mackenroth2011nonlinear}.}
	\label{fig3-1}
\end{figure}

Continued strides in laser technology, especially following the invention of chirped pulse amplification (CPA) have led to the realization, in many laboratories around the world, of ultraintense ($I\sim10^{22}-10^{23}$W/cm$^2$) laser pulses.
The dimensionless parameter $a_0\equiv \lvert e\rvert E_0/(m_ec\omega_0)$ is used to scale the laser field intensity, where $E_0$ and $\omega_0$ are the amplitude and frequency of the laser field, $-e$ and $m_e$ are the electron charge and mass, and $c$ is the speed of light in vacuum.
NLC scattering, driven by such an ultrashort ultraintense laser field, is intuitively understood in Furry picture as follows.
An electron with incoming four-momentum $p^\mu$ enters the laser field, from which it absorbs $n_L$ photons of four-vectors $k^\mu$ and, subsequently, emits a single high-energy photon of four-vector $k'^{\mu}$. Then it continues propagating with a new four-momentum $p'^{\mu}$ until the next radiation event (see Fig.~\ref{fig3-1}).
The QED effects associated with the scattered electron during the NLC process are gauged approximately by the quantum parameter $\chi_e\equiv\lvert e\rvert \hbar\sqrt{(F_{\mu\nu}p^\mu)^2}/m_e^3c^4\chi_e \sim \gamma_e(1-\beta_z)E_0/E_{crit}$, where $F_{\mu\nu}$ is the electromagnetic field tensor, $\hbar$ the reduced Planck constant, $\beta_z$ the longitudinal velocity of electron, and $E_{crit} \approx 1.32\times10^{18}$V/m the critical electric field strength. In the regime of $\chi_e\ll1$, the quantum effects are not pronounced. Under these conditions, the classical T-BMT equations accurately describe the electron spin dynamics in the laser field and the spin flip transitions are almost impossible. Influence of the spin-dependent quantum stocasticity during photon emissions begins to show up in the regime of $\chi_e\sim1$. The electron beam can be rapidly polarized over the femtoseconds time-scale for two important reasons: (1) the electron spin flips asymmetrically, just like in the ST effect in a storage ring \cite{seipt2019ultrafast,song2019spin}; (2) the spin-depended radiation reaction force exerted on the electron (resulting from emission of the high-energy photon) significantly affects the motion of electrons with different spin states \cite{li2019ultrarelativistic} (see Subsect. \ref{sec2-polele}).

\subsection{Polarization-dependent radiation spectra in weakly NLC regime}\label{sec1-polele}
\begin{figure}[!t]
	\setlength{\abovecaptionskip}{0.2cm}  	
	\centering\includegraphics[width=0.9\linewidth]{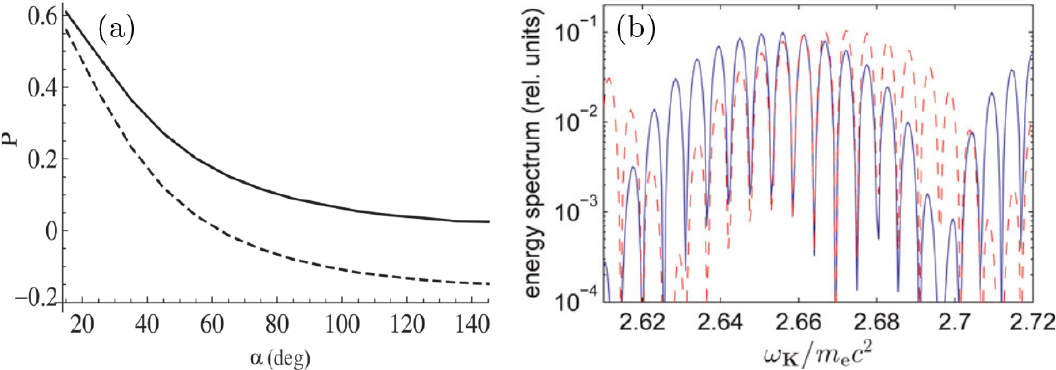}	
	\caption{(a) Polarization of final electron beam($\gamma=5$) as a function of the collision angle \cite{karlovets2011radiative}. (b) The energy spectra of Compton photons emitted in the head-on collision of a short laser pulse and the relativistic electron. Blue-solid curve: the results for the case when the electron helicity is conserved. Red-dashed curve: the results for the case when electron helicity flips \cite{krajewska2013spin}.}
	\label{fig3-2}
\end{figure}

When an ultrarelativistic ($\gamma\gg1$) electron collides with an intense laser with moderate intensity $a_0\gtrsim1$, the scattering process is signified by the weakly NLC scattering due to the limited photon number $n_L\gt 1$ absorbed by electron. The NLC in this regime can be described by the Volkov-state approach or the nonlocal QO approach.
Nikishov and Ritus \cite{nikishov1964quantum,nikishov1967pair} discuss NLC in an external field in the early 1960s. In that work, the probability rate of radiation is obtained from exact solutions to the Dirac and Klein-Gordon equations. Ritus subsequently discussed the spin effects associated with injected electrons in constant crossed fields in the 1970s \cite{ritus1970radiative,ritus1972radiative,ritus1972vacuum}. Bol'shedvorsky \emph{et al.} \cite{bol2000spin} analyze the electron final spin transition as a function of its energy. Ivanov \emph{et al}. \cite{ivanov2004complete} introduce the complete description of spin and polarization effects of weakly NLC in CP and LP laser fields.  The case of an electron colliding head-on with a polarized laser was analyzed in these publications, the scattered electrons possess nonzero polarization only in cases in which polarizations of the initial electrons or laser photons are nonzero.
Karlovets \cite{karlovets2011radiative} considers the exact electron helicity, instead of the Stokes parameters used in \cite{ivanov2004complete}, to describe the polarization of the electrons in the case of non-head-on collisions. In the weakly NLC regime,  the degree of longitudinal polarization can reach about 60\% if the initial electron beam with $\gamma_e\le10$ collides with the CP plane wave at an angle of about 20$^\circ$ [Fig.~\ref{fig3-2} (a)].
In the theoretical investigations cited above, the weakly NLC processes were discussed in constant (crossed) fields or employed monochromatic infinite plane electromagnetic waves.
However, the polarization time required in weakly NLC scattering is much longer than the commonly used picosecond and nanosecond laser pulses, thus rendering the polarization build-up almost impossible. It is, therefore, necessary to discuss NLC scattering in finite plane waves which better model the realistic (laboratory) laser pulses.

In a short laser pulse with $a_0\gtrsim1$, the weakly NLC spectra depend closely on the parameters of the laser configurations employed. The differential or integral radiation spectra, related to arbitrary shape, duration, carrier envelop phase (CEP) and laser polarization, have been extensively investigated \cite{boca2009nonlinear,mackenroth2010determining,mackenroth2011nonlinear,seipt2011nonlinear,krajewska2012compton,dinu2013exact,seipt2013asymmetries,seipt2016analytical,li2018single}, but the electron spin effects have received less attention, overall. Boca \emph{et al.} \cite{boca2012spin} suggest that the electron spin flip should affect the angular distribution of that part of the scattered electron bunch whose energy is much smaller than the energy of the incoming electron beam, i.e., in cases where the electron recoil is large, the spin effects are important.
Krajewska \emph{et al.} \cite{krajewska2013spin,krajewska2014frequency} further discuss the influence of laser pulse duration, CEP, and polarization, on the spin-flip and spin-nonflip Compton-photon and electron spectra. For large Compton-photon energies ($\omega_k\ge2m_ec^2$), the photon spectral intensities of the spin-flip and spin-nonflip  seem to occur with comparable probabilities, and even in some special energy bands, the case of spin-flip [red-solid curve in Fig.~\ref{fig3-2} (b)] dominates over spin nonflip [blue-dashed curve in Fig.~\ref{fig3-2} (b)]. The rapid oscillations depicted in the Compton spectrum of Fig.~\ref{fig3-2} (b) result from multiphoton absorption in NLC. Seipt \emph{et al}. \cite{seipt2011nonlinear} and Krajewska \emph{et al.} \cite{krajewska2013spin,krajewska2014frequency} also discuss the frequency scale between the photon spectra of Thomson and Compton scattering and found out that the spin and polarization are responsible for the main differences.

\subsection{ Polarization build-up of high-energy electrons from
asymmetrically ultraintense laser field}\label{sec2-polele}

In ultraintense laser fields ($a_0\gg1$) the radiation mechanism transits to the strongly NLC regime ($n_L\gg1$) in which the probability of spin-nonflip dominates over that of spin-filp in the case of an incoming electron with low energy (tens of MeV), and the NLC cross section for a Dirac particle is consistent with that of a Klein-Gordon spinless particle \cite{panek2002laser}. Thus the spin effects are still marginal.
However, the situation is different for a high-energy (multi-GeV) incoming electron ($\chi_e\sim1$): the probability of spin-flip is larger than that of spin-nonflip after radiation. In this case,  the spin and polarization modify the Compton scattering and its concomitant radiation-reaction effects, rather significantly  \cite{seipt2018theory,seipt2020spin,li2019ultrarelativistic,li2019electron,chen2022electron,geng2020spin,tang2021radiative}.
The polarization of a high-energy electron in strongly NLC scattering is similar to ST polarization: a TSP electron (positron) beam will be formed spontaneously due to the electron (positron) spin transition to the reverse (same) direction, respectively, as that of the static magnetic field under the influence of synchrotron radiation.

\begin{figure}[!t]
	\setlength{\abovecaptionskip}{0.2cm}  	
	\centering\includegraphics[width=0.9\linewidth]{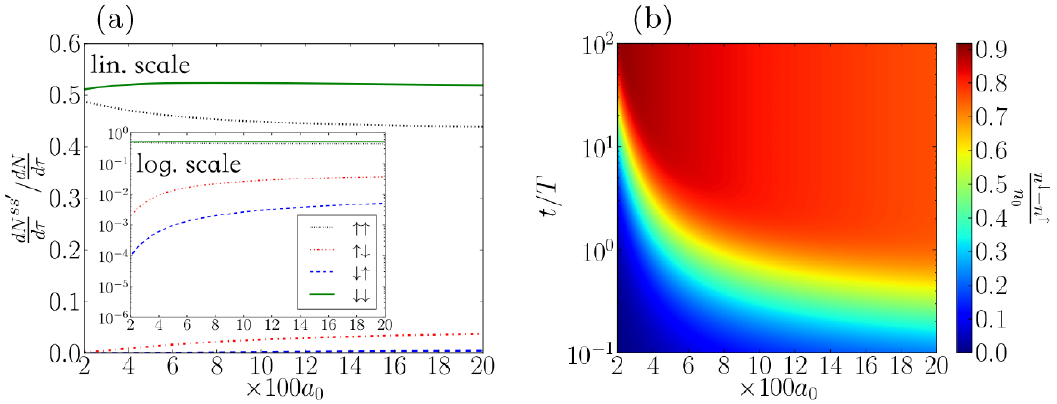}	
	\caption{(a) The spin-flip and spin-nonflip rates, summed over photon polarization and normalized to the unpolarized rate, as function of the strength parameter of the laser electromagnetic waves $a_0$. (b) Degree of electron spin polarization antiparallel as a function of $a_0$ and of time normalized to the laser period $T\approx3.33$ fs \cite{sorbo2017spin}.}
	\label{fig3-3}
\end{figure}
Sorbo \emph{et al}. \cite{sorbo2017spin} derives a simple model to analyse the polarization of electrons in a rotating electromagnetic field, in which the photon emission is described by radiation rates in a constant cross field due to formation length $l_f$ of photon is much small than the laser wave length $\lambda_L$ ($l_f\propto\lambda_L/a_0\ll\lambda_L$). The only nonprecessing spin basis for an electron evolving in the rotating electromagnetic field is
\begin{eqnarray}
	 \hat{\bm\zeta} = \frac{\bm{E}\times \bm{B}}{\Arrowvert\bm{E}\times\bm{B}\Arrowvert},
\end{eqnarray}
each electron in the rotating electromagnetic field has a projection of its spin in the direction of the vector $\hat{\bm\zeta}$ of $\hbar\bm{s}/2$, where $\bm{s} = +1 \equiv \uparrow$ and $\bm{s} = -1 \equiv \downarrow$ mean that electron spin is parallel or anti-parallel to $\hat{\bm\zeta}$, respectively. The photon emission rate $dN^{\uparrow\downarrow}/d\tau + dN^{\downarrow\downarrow}/d\tau$ [red dot-dashed and green solid curve in Fig.~\ref{fig3-3}(a)] is larger than $dN^{\downarrow\uparrow}/d\tau + dN^{\uparrow\uparrow}/d\tau$ [blue dashed and black dot curve in Fig.~\ref{fig3-3}(a)], thus the number of final electrons with $\bm{s}'\equiv\downarrow$ is larger leading to the nonvanished degree of polarization $P = \frac{N^{\downarrow}-N^{\uparrow}}{N^{\downarrow}+N^{\uparrow}}$, as shown in Fig.~\ref{fig3-3}(b).
The time needed to reach about 60$\%$ polarization is about one laser period [see Fig. \ref{fig3-3} (b)], however, regardless of the such short polarization time, electrons are very instable in magnetic nodes \cite{kirk2016radiative}. Sorbo \emph{et al}. \cite{sorbo2018electron} further discuss the spin polarization in the realistic electron trajectory, since the later is subject to the radiation reaction, different initial positions and momenta, and the instability in magnetic nodes. They suggest that the degree of spin polarization reaches an expected value only when timescale for polarization is shorter than that for electrons migrating away from the magnetic node (about 0.01 laser wave length).

\begin{figure}[!t]
	\setlength{\abovecaptionskip}{0.2cm}  	
	\centering\includegraphics[width=0.9\textwidth]{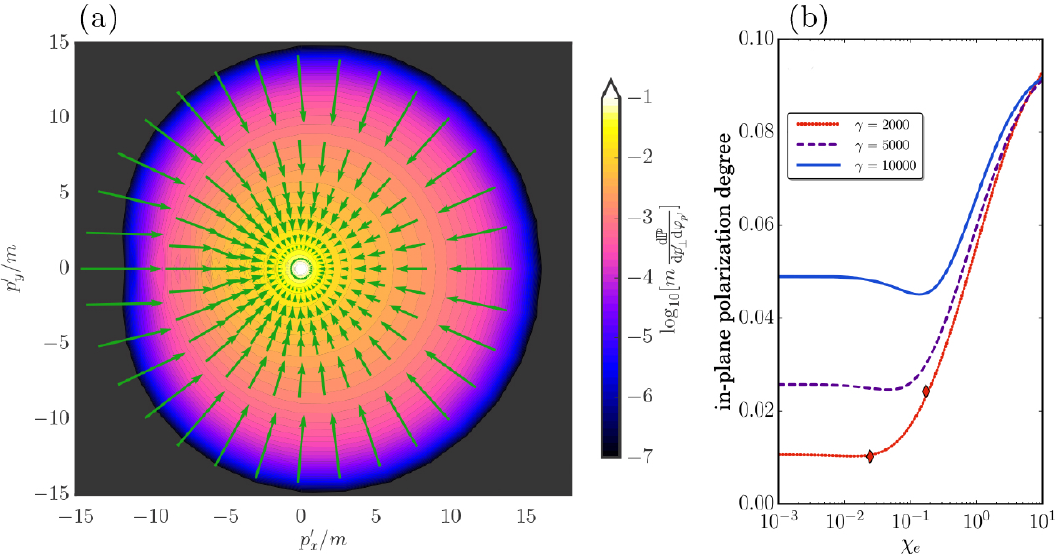}	
		\caption{(a) Transverse momentum distribution of the scattered electrons(as a heat map) and the polarization  of the scattered electron transverse to the beams axis(arrows). (b) degree of polarization in the scattering plane as a function of $\chi_e$ \cite{seipt2018theory}.}
	\label{fig3-4}
\end{figure}

It is convenient to accept the assumption in LCFA that strong fields can be regarded as instantaneously constant within a formation length. In investigations of ultrarelativistic laser interactions with high-energy electrons, the LCFA allows known scattering amplitudes in constant crossed fields (the zero-frequency limit of plane waves) to be adapted to arbitrary fields in simulations, thus facilitating experimental programs \cite{ilderton2019extended}. Spin density matrix schemes \cite{seipt2018theory,seipt2020spin} are well suited for investigating the interaction of an arbitrarily polarized beam with an ultrarelativistic laser field. For example, the degree of radial polarization is nonzero when unpolarized electrons emit only single photons in an ultrashort CP laser pulse, as shown in Fig.~\ref{fig3-4} (a). The dependence of polarization on $\chi_e$ in the scattering plane is shown in Fig.~\ref{fig3-4} (b). It can be inferred that  the degree of polarization reaches up to 0.09 for $\chi_e\simeq10$.

 The NLC radiation formulas in the LCFA regime has been extensively used in PIC simulations for treating the multiple-photon radiation process as a combination of a series of single-photon emissions.  The simulation is semiclassical in which the electron and its spin dynamics between two radiations are described by classical Lorentz and T-BMT equations, respectively, and the photon emission and spin flip are treated by a MC algorithm to calculate the spin-resolved radiation probabilities. Employing Volkov-states, the initial-spin-resolved radiation probability rate and the probability of spin-flip or nonflip after emitting the photons in the LCFA regime are expressed, respectively, as  \cite{seipt2019ultrafast}
\begin{subequations}\label{seiptratio}
\begin{eqnarray}
	\frac{dP}{d\tau} &=& -\frac{\alpha}{b}\int dt\left[ \mathrm{Ai_1}(z) + g\frac{2\mathrm{Ai^{'}}(z)}{z} + \varXi_\zeta\mathrm{sgn}(\dot{h})t\frac{\mathrm{Ai}(z)}{\sqrt{z}}\right], \\\label{ptau}	
	\frac{dP_{\rm nonflip}}{d\tau} &=& -\frac{\alpha}{2b}\int dt \left\lbrace 2\mathrm{Ai}_1 + (g+1)\frac{2\mathrm{Ai}^{'}}{z} + \varXi_\zeta\mathrm{sgn}(\dot{h})\frac{2t-t^2}{1-t}\frac{\mathrm{Ai}}{\sqrt{z}}\right.\nonumber\\
&+&\left.\varXi_\kappa^2\frac{t^2}{1-t}\left[\mathrm{Ai_1}+\frac{\mathrm{Ai^{'}}}{z}\right] \right\rbrace,\label{pnoflip}\\
	\frac{dP_{\rm flip}}{d\tau} &=& -\frac{\alpha}{2b}\int dt \frac{t^2}{(1-t)} \left\lbrace \frac{\mathrm{Ai}^{'}}{z} - \varXi_\zeta\mathrm{sgn}(\dot{h})\frac{\mathrm{Ai}}{\sqrt{z}}\right.\nonumber\\
&-&\left.\varXi_\kappa^2\frac{t^2}{1-t}\left[\mathrm{Ai_1}+\frac{\mathrm{Ai^{'}}}{z}\right] \right\rbrace.\label{pflip}
\end{eqnarray}
\end{subequations}
In these equations $\tau$ is the laser phase, $t$ is the fraction of light-front momentum transferred from the electron to the photon, $g = 1+\frac{t^2}{2(1-t)}$, $b = (\hbar k \cdot p)/m^2c^2$ is the quantum energy parameter, $k$ is the photon four-wavevector, $p$ is the electron four-momentum,  $\mathrm{Ai_1}(z) = \int_{z}^{\infty}\mathrm{Ai}(x)dx$ with $\mathrm{Ai}(x)$ an Airy function, $z = \left[\frac{t}{(1-t)\chi_e}\right]^{2/3}$, and $\varXi_{\kappa,\eta,\zeta}$ denotes components of the initial electron polarization vector along the wave vector, the electric field, and the magnetic field in its rest frame, respectively.
From Eqs. (\ref{pnoflip}) and (\ref{pflip}), any polarization along $\kappa$ direction will decrease the spin-flip rates and increase same the spin-nonflip rates by the equal amount leaving the total rate unchanged. However, the ratios in Eqs. (\ref{pnoflip}) and (\ref{pflip}) linearly depend on $\Xi_\zeta$, the asymmetric flip rates in the direction of $\Xi_\zeta$ means the electrons will polarize along the magnetic field in the rest frame of electron.
\begin{figure}[!t]
	\setlength{\abovecaptionskip}{0.2cm}  	
	\centering\includegraphics[width=0.9\textwidth]{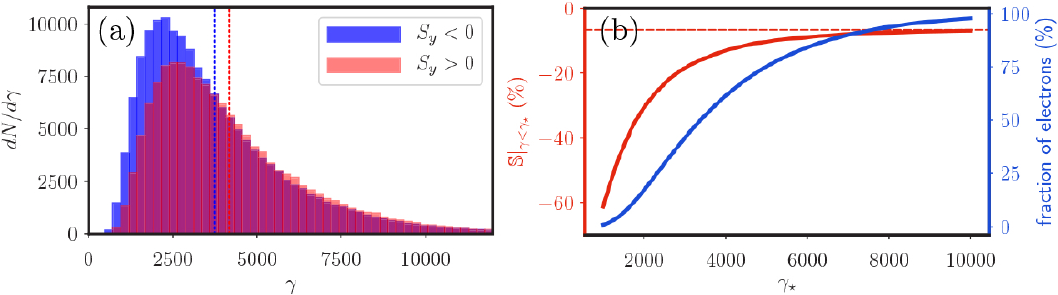}	
    \caption{(a) the final energy distribution for up and down electrons. (b) Polarization of postselected electrons with $\gamma\le\gamma_*$ (left axis) and their relative fraction (right axis) \cite{seipt2019ultrafast}. }
	\label{fig3-5}
\end{figure}
Seipt \emph{et al}. \cite{seipt2019ultrafast} develop a MC algorithm employing the spin-dependent photon emission rates in Eq. (\ref{seiptratio}) and simulate the interaction of an unpolarized electron beam with a bichromatic laser field. The spin polarization leads to a measurable (5\%) splitting of the mean energy of spin-up and spin-down electrons (i.e., spin parallel (antiparallel) to the magnetic field in the rest frame of the electron), as shown in Fig.~\ref{fig3-5} (a). The degree of total polarization is about 6.6\%, but can be increased significantly by post-selecting some electrons, as shown in Fig.~\ref{fig3-5} (b). 

By means of QO approach, the further investigation indicates that the electron beams can be polarized in an asymmetry bichromatic LP laser  field \cite{song2019spin}modeled by $E_x \propto a_1\sin(\omega_L\eta) + a_2\sin(2\omega_L\eta + \phi)$, where $a_1$ and $a_2$ are the invariant laser field parameter and second harmonic pulses, respectively, $\eta = t-z/c$, and $\phi = \frac{\pi}{2}$ is the relative phase. In this scenario, the stronger negative half-cycle results in larger $\chi_e$ than the weaker positive half-cycle. So, the radiation probability in the negative half-cycle will be larger and the spin state probably to flip to the direction of $+y$ after emitting the high-energy photons, will be more, due to $\hat{\bm\zeta}$ being along $-y$. The average polarization of low-energy electrons could reach about 70\%. When asymmetry of the laser strength vanishes, i.e. when $\phi = 0$, the results return back to the LP case in the Ref. \cite{li2018single}, and the net degree of polarization is almost zero. These results are also consistent with ones obtained by Volkov-state approach \cite{seipt2019ultrafast}.

\begin{figure}[!t]
	\setlength{\abovecaptionskip}{0.2cm}  	
	\centering\includegraphics[width=0.9\linewidth]{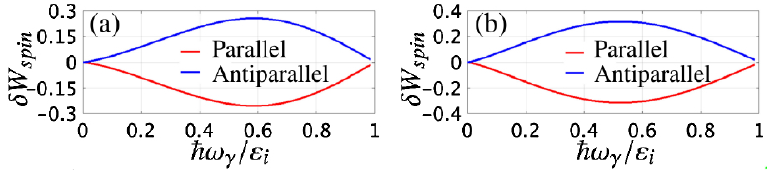}	
	\caption{The relative magnitude of the spin-dependent term in the radiation probability of Eq.(\ref{Wi}) with $\chi_e=1$ (left) and $0.1$ (right), respectively. $\delta W_{spin} \equiv W_{spin} /(W_{rad} - W_{spin})$, and $W_{rad}$ and $W_{spin}$ are the total radiation probability and the spin-dependent term in Eq.(\ref{Wi}). Red and blue curves denote ${\bm{S}}_i$ parallel and antiparallel to SQA,    respectively \cite{li2019ultrarelativistic}.}
	\label{fig3-6}
\end{figure}

In addition to the above mentioned spin-flip mechanism, in the strongly NLC regime, another polarization process of electrons can be induced by the spin-dependent radiation-reaction, which has been clarified by means of QO approach \cite{chen2022electron}. Summing over the final electron spins and photon polarizations, the radiation probability of NLC scattering, depending on the initial spins, is given by \cite{li2019ultrarelativistic}
\begin{eqnarray}
 \frac{{\rm d^2}W_{i}}{{\rm d}u{\rm d}\eta}&=&8{W_R}\left\{-(1+u){\rm IntK}_{\frac{1}{3}}(u')+(2+2u+u^2){\rm K}_{\frac{2}{3}}(u')\right.\nonumber\\
 &-&\left.u{\bm S}_{i} \cdot  \left[{ \bm{\beta}}\times\hat{\bm a}\right]{\rm K}_{\frac{1}{3}}(u') \right\}.
 \label{Wi}
\end{eqnarray}
In this equation, the system of units in which $c = \hbar = 1$ has been used, $W_R={\alpha m}/\left[{8\sqrt{3}\pi \lambda_c \left( k\cdot p_i\right)}{\left(1+u\right)^3}\right]$, $u'=2u/3\chi$, $u=\hbar\omega_\gamma/\left(\varepsilon_i-\hbar\omega_\gamma\right)$, ${\rm IntK}_{\frac{1}{3}}(u')\equiv\int_{u'}^{\infty} {\rm d}z {\rm K}_{\frac{1}{3}}(z)$, $\rm{K}_n$ is an $n$-order modified Bessel function of the second kind,  $\lambda_c$ is the Compton wavelength of the electron, $\omega_\gamma$ is the photon frequency, $\varepsilon_i$ is electron energy before photon emission, $\bm{\beta}={\bm v}/c$ is the velocity scaled by $c$, $\hat{\bm a}={\bm a}/\lvert {\bm a}\rvert$ is a unit vector in the direction of acceleration of the electron, $\eta=k\cdot r$ is the laser phase, $p_i$, $k$, and $r$ denote the electron momentum four-vector before radiation, the laser four-wavevector, and the coordinate four-vector, respectively, and ${\bm S}_{i}$ denotes the electron spin polarization vector before radiation.
The non-precessing spin basis (spin quantization axis, SQA) $\hat{\bm\zeta}= {\bm{\beta}} \times \hat{\bm{a}}$ is along the magnetic field in the rest frame of the electron. According to Eq. (\ref{Wi}),
the emission probability is larger (smaller) when the electron initial spin  ${\bm{S}}_i$ is antiparallel (parallel) to the $\hat{\bm{\zeta}}$. The maximum difference between two states reaches about 60\% for $\chi_e\simeq1$, and about 70\% for $\chi_e\simeq0.1$, see Fig.~\ref{fig3-6}.

\begin{figure}[!t]
	\setlength{\abovecaptionskip}{0.2cm}  	
	\centering\includegraphics[width=0.9\linewidth]{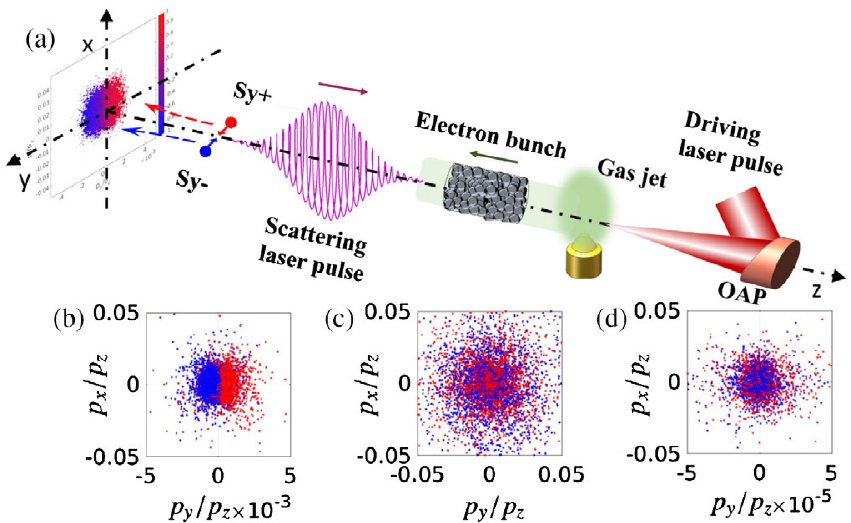}	
	\caption{(a) Scenario of generation of spin-polarized electron beams via NLC scattering. (b), (c) and (d) are the transverse momentum distributions of electrons: for EP, CP and LP laser pulse, respectively. The blue point and red point denote the electrons polarized parallel and antiparallel to the $y-$direction, respectively \cite{li2019ultrarelativistic}. }
	\label{fig3-7}
\end{figure}

The MC simulations indicate that the elliptically polarized (EP) laser has the unique advantage of separating spin-up and spin-down electrons into two groups that propagate along short axis direction of EP laser with a splitting angle of about tens of mrads, see Fig~\ref{fig3-7} \cite{li2019ultrarelativistic}. By contrast, electrons in different spin states still mix after photon-emission in LP and CP lasers. Mechanism of the splitting of polarized electrons induced by an EP laser is explained as follow. In the positive half-cycle ($E_x > 0$) of the elliptically-polarized laser with ellipticity parameter $\epsilon=E_y/E_x = 0.05$, the $y-$component of electron momentum $p_y<0$, and the emission probability of the spin-down electrons, i.e., ${\bm S}_i = -1 \equiv \downarrow$, is larger than that of spin-up electrons, because of $\hat{\bm\zeta} >0$. After emitting the high-energy photons, the spin-down electrons acquire a positive $y-$component of momentum $p'_y>0$, due to radiation-reaction and momentum conservation. Conversely, in the negative half-cycle of the laser field ($E_x < 0$), $\hat{\bm\zeta} < 0$, spin-up electrons emit photons with higher probability, and acquire nagetive $y-$components of momentum $p'_y<0$. Finally, the opposite spin state would be separated into two parts with a splitting angle of about 20 mrad, with the average polarization of each part reaching 70\%. It should be borne in mind that the electrons deflect due to the spin-dependent radiation-reaction force during NLC scattering, rather than due to the Stern-Gerlach force, because the latter is much smaller than the former and the deflection angle is only about $10^{-4}$ mrad in the collision of 10 MeV electrons with a $10^{22}$ W/cm$^2$ laser pulse \cite{wen2017spin}. The spin-dependent stochasticity in photon-emission processes, taking into account the radiation-reaction effects, have also been investigated in detail \cite{guo2020stochasticity,geng2020spin}. Furthermore, the loop spin effects are also figured out in extremely strong fields, where electrons may polarize without radiating via electromagnetic self-interaction \cite{meuren2011quantum,ilderton2020loop,torgrimsson2021loops}.

In the MC simulation algorithm of strongly NLC, after emitting photons, the spin state of electron collapses onto one of its basis states defined with respect to the instantaneous SQA (defined in \cite{li2019ultrarelativistic}). This method has been successfully employed in the unpolarized and transversely polarized electrons and has later been extended
to arbitrary initial spin configurations in Ref. \cite{xue2021generation}; Besides, alternatively  similar approaches are proposed in Refs. \cite{geng2019generalizing} and \cite{tang2021radiative}, which have successfully reproduced the results in Ref. \cite{li2018single}.

%which losts the complete 3D information on the spin polarization. A general ST polarization is described through the spin-flip rate in Eq. (\ref{pflip}) \cite{geng2019generalizing}, by this model the evolution of full spin formation is discussed in strongly NLC, the results show that the SQA used in \cite{li2019ultrarelativistic} is applicable for the cases of unpolarized and transversely polarized incoming electron beam. Moreover, the entire spin information is discussed by calculating the change of the spin polarization vector $\Delta \varXi_{\kappa,\eta,\zeta}$ after the emission in complicated laser field, as well as the spin transition in the case of no photon emission \cite{tang2021radiative}, which improves the MC simulation to maintain the complete information on spin polarization of the electron at any simulation moment.
\section{Polarization build-up of $\gamma$-photons in nonlinear Compton scattering}\label{sec-polgamm}

A plasma-wakefield electron beam generated by an ultraintense laser  can be used to build a high-brilliance, ultrashort-pulse, $\gamma$-photon beam via collision with a scattering intense laser. A quasi-monoenergetic MeV $\gamma$-photon beam with brilliance above $\sim 3\times10^{20}$ photons /(s mm$^2$ mrad$^2$ 0.1\% BW) can be produced in experiments, via an all-optical setup \cite{phuoc2012all,chen2013mev,liu2014generation,sarri2014ultrahigh,yu2016ultrahigh,albert2016applications,yan2017high}. Production of LP and CP photons with $\sim10^7$ yield in a single shot using a compact all-optical, polarized, X-ray source based on inverse Compton scattering, has been reported recently \cite{ma2022compact}. Utilizing the polarized electron beams in a plasma wakefield, proposed in Sect. \ref{sec-polacc}, a high-brilliance, polarized, $\gamma$-photon beam can possibly be generated in the future.
In this section, we review the production of polarized $\gamma$ photons in two regimes. In one regime, polarization is from the laser photons in weakly NLC processes, and in the second regime, polarization is from electrons in strongly NLC scattering. Work investigating the transfer of orbital angular momentum between the emitted $\gamma$ photons and the laser-photon angular momentum will also be reviewed.

\subsection{$\gamma-$ photon polarization from laser-photons in the weakly NLC scattering regime}\label{sec1-polgamm}

\begin{figure}[!t]	
	\setlength{\abovecaptionskip}{0.2cm}  	
	\centering\includegraphics[width=0.8\linewidth]{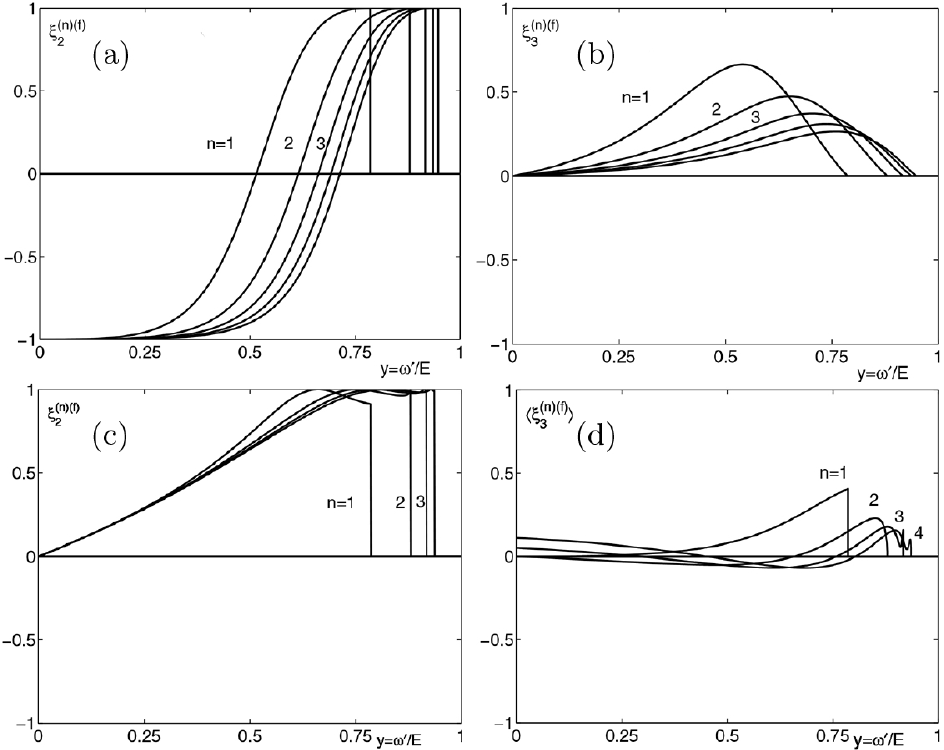}
	\caption{The Stokes parameters of the final photons for different harmonics $n$ versus the final photon energy $\omega'$ at $\zeta_3 = 1$ (initial electron helicity), $a_0^2=0.3$ with: (a)-(b) CP laser photons, (c)-(d) LP laser photons. $\xi_2^{(n)(f)}$: circular polarization; $\langle\xi_3^{(n)(f)}\rangle$: linear polarization along a basis vector averaged over azimuthal angle \cite{ivanov2005complete}. }
	\label{fig4-1}
\end{figure}

The weakly NLC regime is similar to the linear Compton process in that polarization of the emitted $\gamma$-photons is determined by polarization of the driving laser, due to the fact that the radiation formation length is much larger than the laser wavelength.
A complete description of the NLC scattering in the case of CP and LP laser photons, in which polarization of all the other particles is arbitrary, is provided by the theory of Nikishov and Ritus \cite{ivanov2005complete}. When an electron with positive helicity absorbs CP or LP laser photons, the energy-dependent mean helicities, $\xi_2^{(n)(f)}$, of the final photons can be presented by the few first harmonics, and each harmonic is 100\% polarized near its maximum $y = y_n$ [see Figs. \ref{fig4-1} (a) and (c)]. The interesting feature related to the linear polarization, $\xi_3^{(n)(f)}$, of the final photons is that the Stokes parameter averaged over the azimuthal angle vanishes [see Fig. \ref{fig4-1} (b) and (d)].

\begin{figure}[!t]	
	\setlength{\abovecaptionskip}{0.2cm}  	
	\centering
	\includegraphics[width=0.8\linewidth]{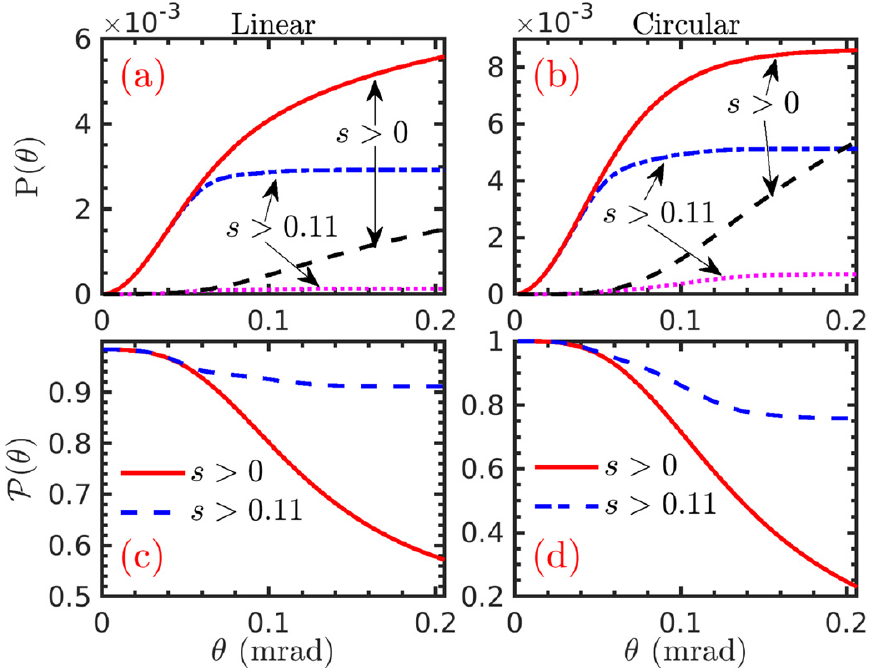}
	\caption{Photon production probability $P$ and polarisation degree $\mathcal{P}$ for different detector angular acceptance values, $\theta$. Left column: linearly polarised background. Right column: circularly polarised background. $s$ is the lightfront momentum fraction of the scattered photon \cite{tang2020highly}.}
	\label{fig4-2}
\end{figure}

 In a head-on collision between a GeV electron and an ultrashort laser pulse with moderate intensity,  the energy-dependent circular polarization is obtained by exact Volkov-state approach and illustrated by ones obtained from harmonic results (the perturbative treatment) \cite{king2020nonlinear}. Calculations within the context of this scenario of the angular and spin-resolved differential probabilities \cite{tang2020highly} indicate that the total number of detectable photons decreases with a narrower acceptance angle, and the degree of polarization increases sharply by narrowing the detector acceptance angle (see Fig. \ref{fig4-2}). A robust scheme to produce a polarized GeV-photon source is proposed, with polarisation degrees exceeding 91\% and 78\% (corresponding to a 96\% and 89\% fraction of the photons) for LP and CP $\gamma$ photons, respectively and with a brilliance of the order of $10^{21}$ photons/(s mm$^2$ mrad$^2$ 0.1\% BW).
 The QO approach is also improved by producing numerically feasible expressions, which can be used to perform NLC scattering calculations. From such calculations, employing moderate laser intensities, the obtained spectral structures and polarization of the radiated photons are in agreement with the results obtained from the Volkov-state approach in the case of a background plane wave  \cite{wistisen2014interference,wistisen2019numerical}.

\subsection{$\gamma-$photon polarization from electrons in the strongly NLC scattering regime}\label{sec2-polgamm}

When the path of an ultrarelativistic particle crosses the propagation direction of an ultraintense laser beam, the total angle of particle deflection in the field is large compared to the radiation cone angle $1/\gamma$. Then the formation length is negligible compared with the length of the field inhomogeneity. Thus the polarization characteristics of the $\gamma$-photons depend mainly on the electron spin. In this sense, the emission process resembles bremsstrahlung.
\begin{figure}[!t]	
	\setlength{\abovecaptionskip}{0.2cm}  	
	\centering
	\includegraphics[width=0.8\linewidth]{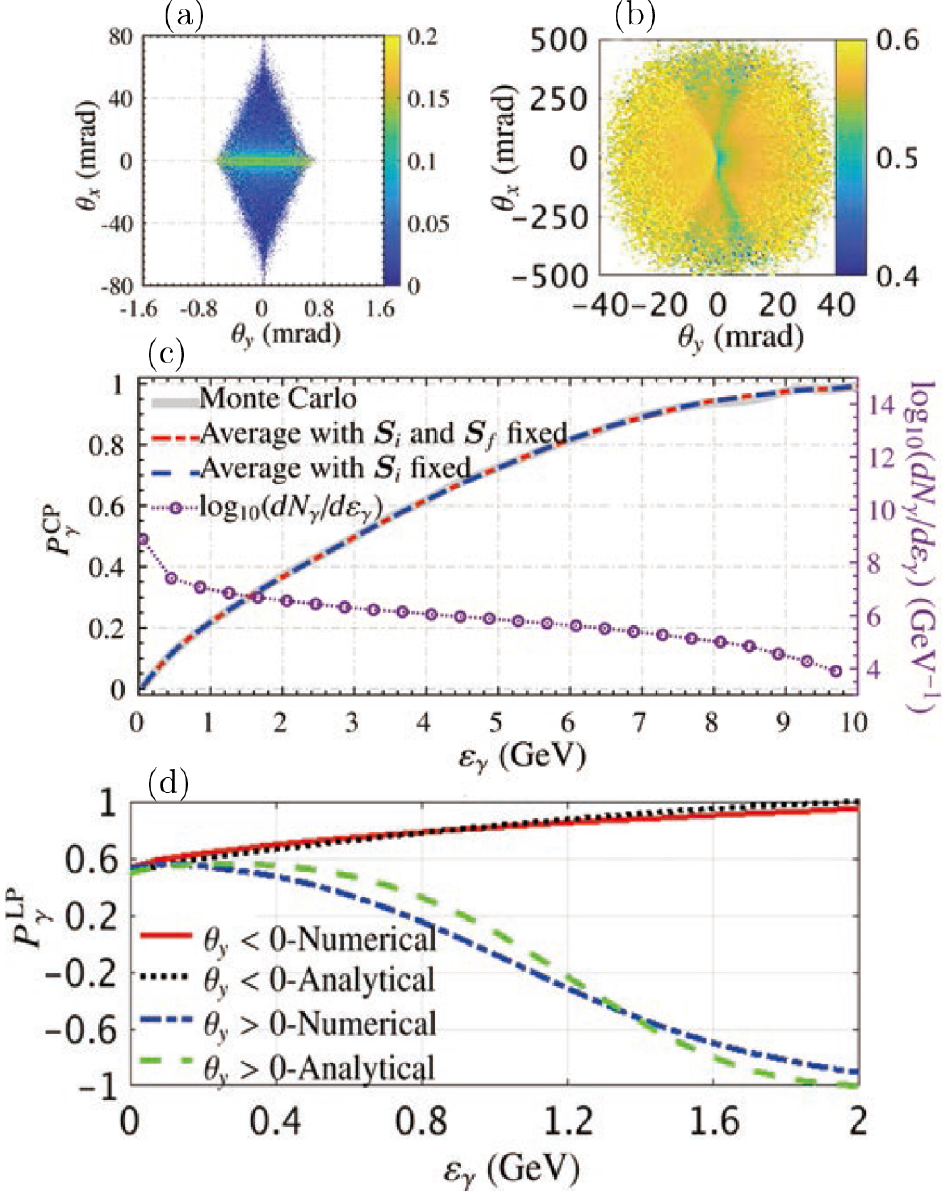}
	\caption{(a) Distribution of average circular polarization for all emitted photons $\overline{P^{CP}}=\xi_2$; (b) Average linear polarization for all emitted photons $\overline{P^{LP}}=\xi_3$ vs $\theta_x$ and $\theta_y$; (c) Degree of circular polarization of $\gamma$-photons $P^{CP}_\gamma=\xi_2$ and energy density of $\gamma$ photons vs $\gamma$-photon energy $\epsilon_\gamma$; (d) Linear polarization of $\gamma$ photons $P^{LP}_\gamma$ (employing $\xi_3$) vs $\varepsilon_\gamma$ \cite{li2020polarized}.}
	\label{fig4-4}
\end{figure}

Production of CP and LP $\gamma$-photons via the single-shot interaction of an ultraintense laser pulse with a spin-polarized counter-propagating ultrarelativistic electron beam has been investigated in the quantum radiation-dominated regime described by the QO approach \cite{li2020polarized}. In the scheme of an arbitrarily polarized laser pulse colliding with a LSP electron bunch, the circular polarization of the produced $\gamma$-photons is transferred from the angular momentum (helicity) of the electrons [see Fig.~\ref{fig4-4} (a)]. The larger the average energy of the emitted $\gamma$-photons, the smaller the number of emitted photons per electron. Thus the transferred average helicity per photon will be larger in the case of high photon energies [see Fig.~\ref{fig4-4} (c)]. For instance, the produced multi-GeV $\gamma$-photons can have circular polarization above 94\% and high brilliance comparable to the unpolarized multi-MeV $\gamma$-photons produced  in experiments  \cite{sarri2014ultrahigh}. It is also found that an EP laser pulse colliding with a TSP electron bunch can produce a higher degree of linear polarization of $\gamma$-photons than that from an LP laser,  and the average degree of linear polarization can reach about 58.3\% [see Fig.~\ref{fig4-4} (b) and (d)].

\begin{figure}[!t]	
	\setlength{\abovecaptionskip}{0.2cm}  	
	\centering
	\includegraphics[width=0.8\linewidth]{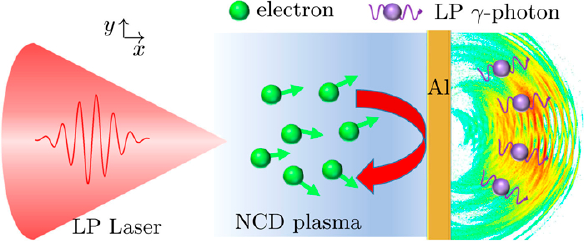}
	\caption{Scenario for the generation of LP $\gamma$-rays via NLC scattering \cite{xue2020generation}. }
	\label{fig4-5}
\end{figure}

 Production of highly LP high-energy brilliant $\gamma$-photons via laser-plasma interactions has been investigated in 3D PIC simulations, coupled with the strong-field QED processes \cite{xue2020generation}. When an ultraintense LP laser pulse irradiates a near-critical-density (NCD) plasma followed by an ultrathin planar aluminum target, the electrons in the NCD plasma are first accelerated by the driving laser to ultrarelativistic energies and then collide head-on with the laser pulse reflected by the aluminum target (see Fig.~\ref{fig4-5}). The end result is emission of brilliant LP $\gamma$-photons with an average degree of polarization  $\sim$70\% and energy up to hundreds of MeV.

In addition to playing a central role in the design for polarized $\gamma$-photon sources, the polarized NLC processes also couple with the NBW processes (the strong-field QED cascades) during the interaction of ultraintense laser with an electron beam or solid target. Thus the polarization effects of abundant $\gamma$-photons are non-negligible in laser-driven QED cascades, especially on the yields of BW pairs.
Considering the photon-seeded pair-creation for LP photons in constant crossed fields, simulations have shown that the approximation of using unpolarized cross-sections for tree-level processes predicts the same number of produced particles when using polarized cross sections to within around 5\%. On the other hand, the simulations showed that when the polarization of the photon can be influenced by its environment, the asymmetry in the polarization distribution could be used to significantly increase the rates of each process \cite{king2013photon}. Evolution of the polarized QED cascades, which takes the spin and polarization of the electrons, positrons and photons into account, is modeled by a Boltzmann-type kinetic equation \cite{seipt2021polarized}. The simulated results show that although the growth rate of QED cascades in ultra-intense laser fields can be substantially reduced, the produced particles are highly polarized.

It has recently pointed out that the photon polarization can significantly affect the pair yield by a factor of more than 10\%, and the considered signature of the photon polarization in the pair yield can be experimentally identified in a prospective two-stage setup \cite{wan2020high}.
By means of QED-PIC simulations, the polarized QED cascades are also investigated through the interaction of two counter-propagating LP laser pulses of peak intensity $8.9\times10^{23}$ W cm$^{-2}$ with a thin foil target \cite{song2021spin}. The simulation results show that the average degree of linear polarization of the emitted $\gamma$-photons can exceed 50\%, which leads to a reduced positron yield by about 10\% (such results are consistent with those in Refs. \cite{xue2020generation,wan2020high}). Additionally, it is pointed out that the pair productions in the linear BW process can dominate over ones in the NBW process when a laser pulse with a peak intensity $\gtrsim10^{22}$ Wcm$^{-2}$ propagates through an overdense plasma channel. Such an interaction scenario induces the relativistic transparency and the coupled NLC, NBW and linear BW processes \cite{he2021dominance,he2021a}. Therefore, the effects of $\gamma$-photon polarization could be no longer negligible in the dynamics of QED plasmas as the yield and kinematics of linear BW pairs have remarkable dependence on the photon polarization \cite{zhao2022signatures}. Furthermore, it is also pointed out that under the moderate laser intensity,  the high-order nonlinear QED processes can be approximated by the cascades between the first-order NLC and NBW processes \cite{dinu2020approximation}.

\begin{figure}[!t]	
	\setlength{\abovecaptionskip}{0.2cm}  	
	\centering
	\includegraphics[width=0.8\linewidth]{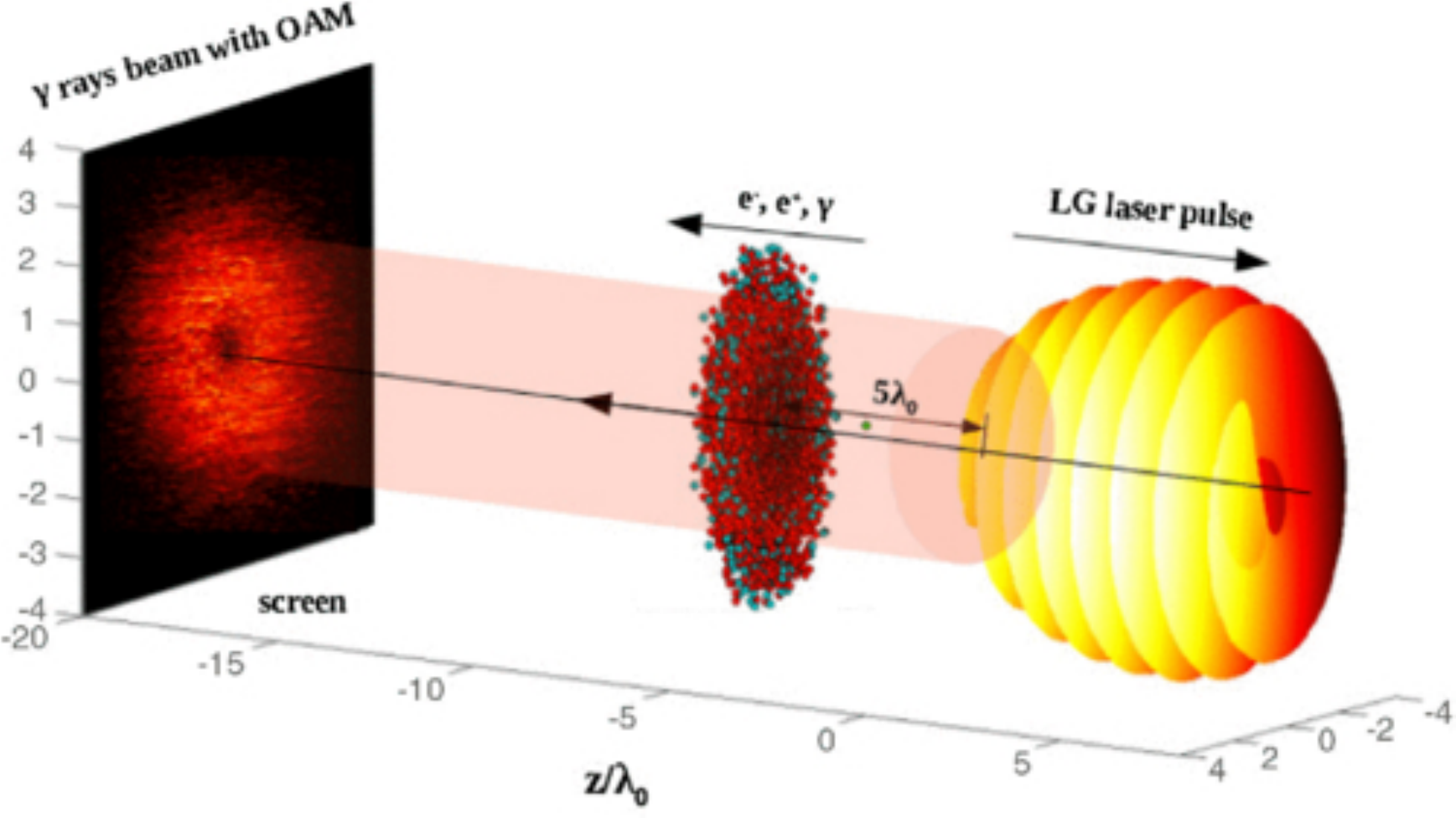}
	\caption{Scheme for generation of a $\gamma$-ray beam with OAM.  An intense Laguerre-Gaussian laser beam of linear (illustrated in the figure) or circular polarization counterpropagates with and scatters off an electron beam \cite{chen2018gamma}.}
	\label{fig4-6}
\end{figure}

Besides the spin angular momentum (SAM), $\gamma$-photon beams with a large orbital angular momentum (OAM) have potential applications in nuclear physics, quantum entanglement and astrophysics \cite{fickler2012quantum,tamburini2011twisting,wang2012terabit,taira2017gamma}. In the classical regime, the production of OAM photons is described by Thomson scattering using a CP laser or an helical undulator \cite{taira2018gamma}. The semiclassical probability of radiation of twisted photons has been obtained to describe the interaction of ultrarelativistic particles moving in the electromagnetic field \cite{bogdanov2019semiclassical}. Based on the strongly NLC scattering, recent, numerous studies have proposed the production of OAM $\gamma$-photons using a Laguerre-Gaussian (LG) laser beam, since LG-laser photons carry spin and orbital angular momenta (SAM and OAM, respectively) \cite{chen2018gamma,ju2019generation,feng2019the,liu2020vortex,wang2020generation}.
Numerical simulations demonstrate the angular momentum conservation among absorbed LG-laser photons, quantum radiation, and electrons in the quantum radiation-dominated regime \cite{chen2018gamma}, and show that the angular momentum of the absorbed laser photons is not solely transferred to the emitted $\gamma$-photons, but is shared between the $\gamma$-photons and the interacting electrons, because of the radiation-reaction effects (see Fig. \ref{fig4-6}). By 3D PIC simulation of LG laser-pulse interactions with an underdense plasma, the production of MeV $\gamma$-photons with high brilliance, a small divergence angle, and controllable angular momentum, has recently been proposed \cite{ju2019generation}. In such a process, the $\gamma$-photon beam acquires angular momentum from the helically distributed relativistic electrons driven by the LG-laser and the self-generated fields in the plasma bubble. Moreover, 3D-PIC simulations reveal that the SAM of the CP laser can be transferred to OAM of the accelerated electrons and further to the emitted $\gamma$-photons, which is achieved by resonant acceleration of electrons in CP laser-plasma interactions \cite{feng2019the,wang2020generation}.

%Three-dimensional particle-in-cell simulations show that vortex $\gamma$-ray beams with a large and controllable OAM of 2.5$\times$ 10$^{18}\hbar$ and energy up to hundreds of MeV can be obtained by using two lasers at intensities of approximately 10$^{22}$ W cm$^{-2}$, which is much more practical and more efficient than previously proposed NLC of a LG beam.

\section{Production of a polarized positron beam driven by ultraintense laser}\label{sec-polposi}
The NBW electron-positron pair production, i.e., $\gamma + n_L\omega \rightarrow e^- + e^+$  ($\omega$ is coherent laser-photon frequency),  is the cross-symmetry process of NLC in Fig. \ref{fig3-1}.
The NBW process in the interaction of an electron beam, of energy in the tens-of-GeV range, colliding with a laser pulse of $a_0\lesssim1$, has been verified in the famous E144 operated at the SLAC facility \cite{Bula1996Observation,Bamber1999Studies,Burke1997Positron}. Further advances in petawatt laser technology have stimulated much theoretical activity in the field, leading to several proposals for the laser-driven production of polarized positrons  \cite{Buscher2020Generation,chen2019polarized,wan2020ultrarelativistic,li2020production,xue2021generation,song2021dense,liu2020trapping}. In most of these proposals, high-energy $\gamma$-photons are emitted in NLC scatting during the interaction of an ultraintense laser with an ultrarelativity electron beam or a solid target, with the $\gamma$-photons continuing to propagate through the laser to produce $e^+e^-$ pairs. Ultraintense laser-driven pair production display the characteristics of high conversion, ultrashort pulse and dense current. Work that investigates the spin and polarization effects in the NBW process, and some typical proposals for polarized positron sources, will be reviewd below.

\subsection{Spin-polarization build-up in the nonlinear Breit-Wheeler process}\label{sec1-polpos}
\begin{figure}[!t]
	\centering
	\includegraphics[width=0.8\linewidth]{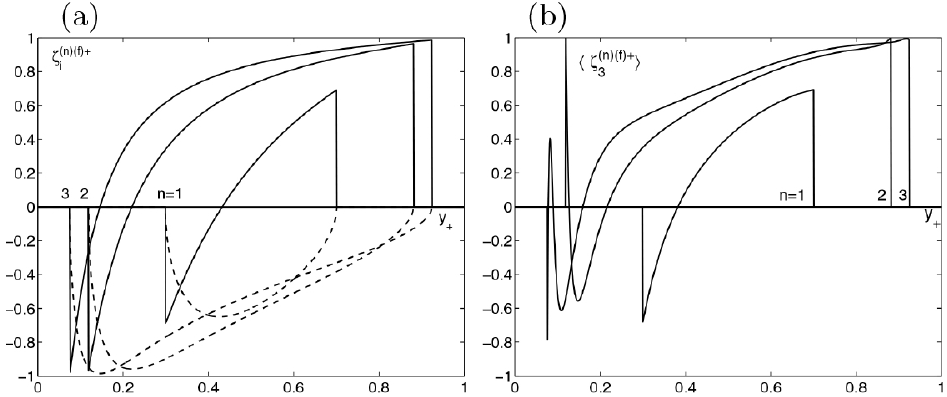}
	\caption{Polarization of positrons for different harmonics $n$ at
		$a_0^2 = 0.3$. (a)  The laser and $\gamma$ photons are CP. Solid curves: helicity $\zeta_3^{(n)(f)+}$, dashed curves: transverse polarization $\zeta_2^{(n)(f)+}$. (b) The laser and $\gamma$ photons are LP and CP, respectively. $\langle\zeta_3^{(n)(f)+}\rangle$: averaged longitudinal polarization  over azimuthal angle \cite{ivanov2005complete}.  }
	\label{fig5-1}
\end{figure}
Theoretical description of the effects of spin and polarization in NBW process  have long been widely studied under various configures of electromagnetic fields.
In a monochromatic plane wave, Ritus and Nikishov  \cite{nikishov1964quantum,ritus1972vacuum,ritus1985quantum} study the effects of photon polarization, and give the probabilities of pair formation by an unpolarized and arbitrarily polarized photon. Ritus \cite{ritus1970radiative} also discuss the radiative effects of the NLC scattering and the NBW process at $\chi_{\gamma,e} \gtrsim 1$ in a constant-crossed field and find that the square of the photon mass and the change of the electron mass increase like $\chi_{\gamma,e}^{2/3}$. Tsai \textit{et al.} \cite{Tsai1974Photon} and Katkov \textit{et al.} \cite{Katkov2012Production} study the pair production of polarized photons in a constant magnetic field via the proper-time method presented by Schwinger and arbitrarily-constant electromagnetic field by analysing the imaginary part of the diagonalized polarization operator, respectively. Furthermore, the effects of photon polarization on the electron-seeded pair-creation cascades under constant cross field, and on the pair yield of ultraintense laser-driven NBW have been investigated in numerical simulations (see. Subsect. \ref{sec2-polgamm}).

The spin effects related to pair-production in the nonlinear BW process have attracted more recent attention. Tsai \cite{Tsai1993Laser} discuss the helicity amplitudes of interaction of laser-electron scattering and $laser+ \gamma \rightarrow e^+ + e^-$ in a monochromatic plane wave, and suggested mechanisms to produce nearly 100\% circularly-polarized $\gamma$ and $e^\pm$ beams. Ivanov \emph{et al}. \cite{ivanov2005complete} give a complete description of the polarization effects in an NBW process by CP or LP laser photons. For a CP $\gamma$-photon, whether it interacts with a CP or LP laser, the produced positrons in the high-energy tail of the energy spectrum have highly longitudinal polarization, like in the transfer of electron spin to photon polarization in NLC (see Fig.~\ref{fig5-1}). By using a density matrix in the Volkov-state approach with LCFA, Seipt \textit{et al.} \cite{seipt2020spin} obtain the completely polarized probability rates of NLC scattering and the NBW process, with arbitrarily polarized photons and electrons. The polarized NLC and NBW processes in a plane-wave laser pulse of arbitrary temporal shape can both be described in this approach, and are also employed to study the polarized QED cascades (see Subsect. \ref{sec2-polgamm}) \cite{seipt2021polarized}. By using a density matrix to describe the particle (electron and positron) spin and photon polarization, Tang \textit{et al.} \cite{tang2022fully} obtain the completely polarized probability rates for the NBW process beyond the LCFA regime in the Volkov-state approach, and investigate the fully polarized NBW process in a pulsed plane wave background. Simulation results suggest that the positron yield can be improved by orthogonalizing the photon polarization to the laser polarization (which has also been demonstrated in Ref. \cite{wan2020high}). Based on the QO approach, Wistisen \cite{wistisen2020numerical} also obtains the numerically feasible probability ratios for the NBW process, which can be used for a laser pulse with arbitrary intensity, and Chen \cite{chen2022electron} \emph{et al.} obtain the fully polarization-resolved probabilities for strong-field QED processes within the LCFA regime. 

The spin and polarization effects in the NBW have also been discussed in other types of strong-field backgrounds, such as in combined Coulomb and strong laser fields \cite{piazza2010polarization}, in a rotating electric field \cite{kohlfurst2019spin}, and in a strong slow electric field superimposed by a weak fast electric field \cite{huang2019spin}.

  \begin{figure}[!t]
	\centering
	\includegraphics[width=0.8\linewidth]{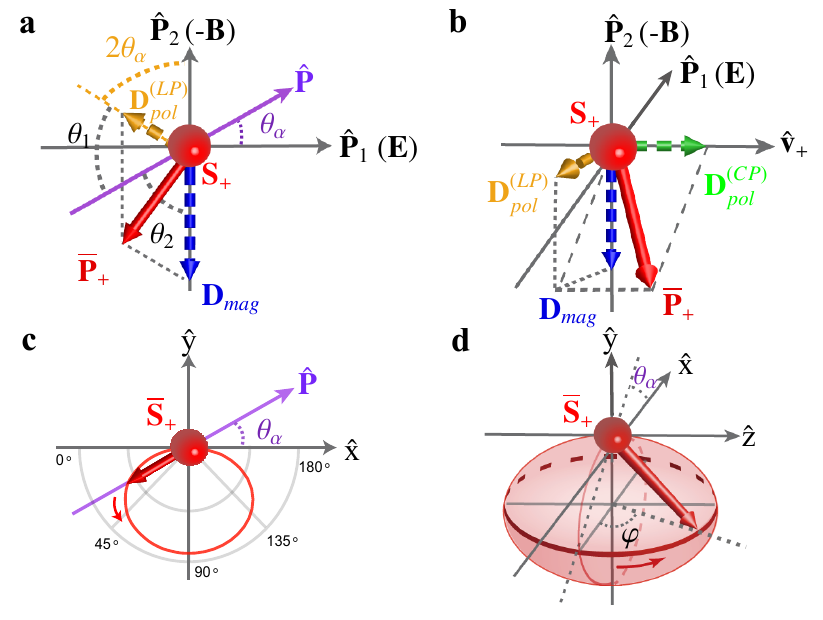}
	\caption{ { The polarization mechanism of arbitrarily polarized positron beams. } (a) and (b): Spin of a sample positron ${\mathbf S}_+$ created by LP (a) and EP (b) $\gamma$-photons, respectively.
		${\overline{\mathbf P}}_{+}$ indicates  the average polarization vector (i.e. the instantaneous SQA) with two transverse components ${\mathbf D}_{mag}$ and ${\mathbf D}_{pol}^{(LP)}$ and one longitudinal component ${\mathbf D}_{pol}^{(CP)}$ (see the analytical formula in Eq.~8 \cite{xue2021generation}).
		For LP $\gamma$-photon in (a), ${\mathbf D}_{pol}^{(CP)}=0$ and the positron polarization is engineered in the transverse plane, while in the case of EP $\gamma$-photon in (b), 3-dimensional control of ${\overline{\mathbf P}}_{+}$ is possible. {Compared with the case of a LP photon, the polarization $\hat{{\mathbf P}}$ of an EP photon is less intuitive and thus not shown in (b).}
		In our interaction scheme, $\hat{\mathbf P}_1 = \hat{\mathbf E} \approx \hat{{\mathbf a}}_+$ and $\hat{\mathbf P}_2 = -\hat{\mathbf B} \approx \hat{{\mathbf b}}_+$; $\theta_\alpha$ indicates the polarization angle to $\hat{\mathbf P}_1$; $\theta_1$ and $\theta_2$ are the angles of $\hat{\mathbf P}$ to ${\mathbf D}_{pol}^{(LP)}$ and ${\mathbf D}_{mag}$, respectively.
		(c) and (d):  Averaged-over-energy polarization
		of the positrons $\overline{\mathbf S}_+$ created by LP (c) and EP (d) $\gamma$-photons, respectively. The red arrow and circle (ellipsoid) indicate the direction and amplitude of $\overline{\mathbf S}_+$ (see the analytical formula in Eq.~1 \cite{xue2021generation}), respectively.}\label{fig5-2}
\end{figure}

Based on the fully spin-resolved QO approach, Xue \emph{et al.} \cite{xue2021generation} use the fully polarization-resolved MC simulation to clarify the effects of photon polarization and asymmetry of the ultraintense laser field on the spin of the pair in the NBW process. Due to the asymmetric spin-dependent pair-production probabilities, the positrons are polarized at the moment of generation, but since the positron polarization direction depends on the direction of the field, positrons will not be polarized in a symmetrical laser field. They suggested a numerical simulation method to estimate the polarization of positrons created by polarized photons in an asymmetric laser field, based on
\begin{eqnarray}
	\overline{{\mathbf S}}_+ =\left[ {\cal A}_{field}\cdot \left(\overline{\mathbf D}_{mag}+\overline{\mathbf D}_{pol}^{(LP)}\right)+\overline{\mathbf D}_{pol}^{(CP)} \right] \big/{\overline{C}},
	\label{estimation1}
\end{eqnarray}
where $\overline{\mathbf D}_{mag}$ indicates the effects of the laser magnetic field on the positron polarization, $\overline{\mathbf D}_{pol}^{(LP)}$ and $\overline{\mathbf D}_{pol}^{(CP)}$ denote the spin angular momentum transfer from linear and circular polarization components of polarized photons to transverse and  longitudinal components of positron polarization, respectively, and ${\cal A}_{field}$ is the laser asymmetry (see a image display in Fig.~\ref{fig5-2}). Dai {\textit{et al.}} \cite{dai2022photon} also investigate the correlated photon and electron (positron) polarization effects with a similar algorithm and simulate the generation of polarized positron in the head-on colliding of a transversely polarized electron beam with an elliptically polarized laser beam.

\begin{figure}[!t]
	\centering
	\includegraphics[width=0.8\linewidth]{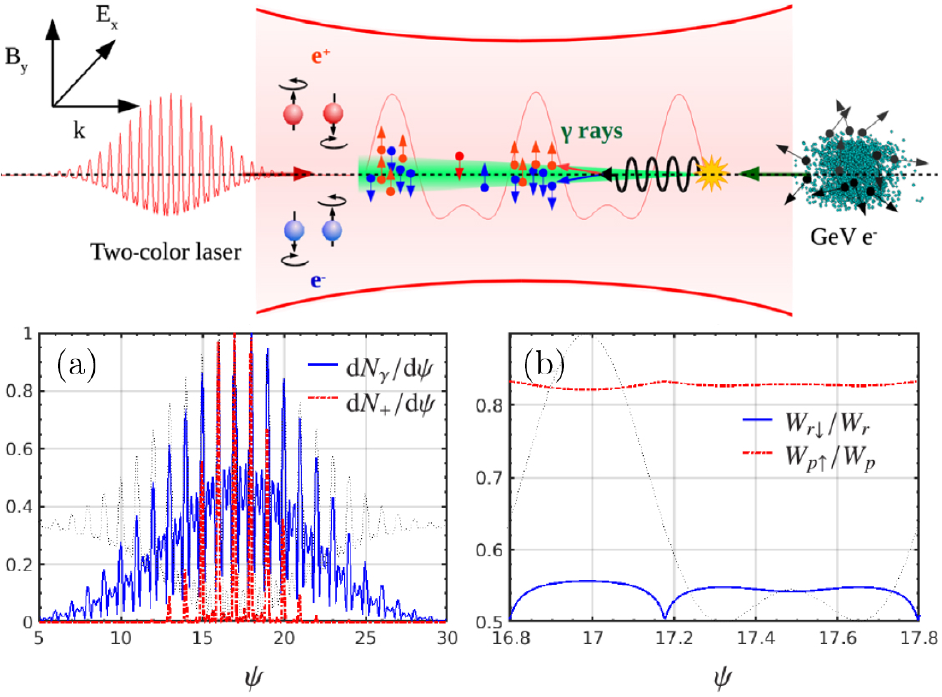}
	\caption{Scheme of laser-based polarized positron beam production\cite{chen2019polarized}. (a) Photon number $dN_\gamma/d\psi$ emitted by seed electrons $e_0^-$(solid blue line) at emission points and positron number $dN_+/d\psi$ (dashed red line) at production points. (b) Ratio of electron radiation probability with spin antiparallel to the magnetic field direction (solid blue line), and ratio of pair production probability with the positron spin parallel to magnetic field direction (dashed-dotted red line) \cite{chen2019polarized}.}\label{fig5-3}
\end{figure}
In addition to the effects of photon polarization and pair spin, the laser parameters also play a significant role in the pair-production process. The laser envelope has obvious modulation on the light scattering cross-section when photons interact with a polarized laser \cite{Titov2012Enhanced}. In a CP laser, interference on the length scale of the pulse envelope, and smoothness of the pulse edges, are found to influence the pair spectrum \cite{Tang2021Pulse}. The yield of pair-production is also related to the polarization of the laser and is very sensitive to changes in the chirp parameters \cite{Obulkasim2019Pair}.

\subsection{Proposals for the production of polarized positrons by an ultraintense laser}\label{sec2-polpos}

Building on the progress made in the production of high-energy $\gamma$-photons in the strongly NLC scattering (reviewed in Sect. \ref{sec-polgamm}), some schemes have been proposed for the production of high-energy polarized positron beams in NBW processes \cite{chen2019polarized,li2020production,xue2021generation}. According to Eq.(\ref{estimation1}), the produced positrons acquire polarization from two sources: i) asymmetry of the intensity of the laser field, i.e. ${\cal A}_{field}\neq 0$, and ii) circular polarization of the intermediate photons, i.e. $\overline{\mathbf D}_{pol}^{(CP)} \neq 0$, and then they decay into electrons and positrons with longitudinal polarization via helicity transfer.
Besides, the positrons produced in different spin states would be separated in space by interaction with the asymmetrically EP laser-beam or as a result of the laser-solid interaction driven by the ultraintense laser (to be referred to, henceforth, as {\it spin splitting}), which leads to the emergence of highly polarized positron beams through selected angles. \cite{wan2020ultrarelativistic,song2021dense}.

\begin{figure}[!t]
	\centering
	\includegraphics[width=0.8\linewidth]{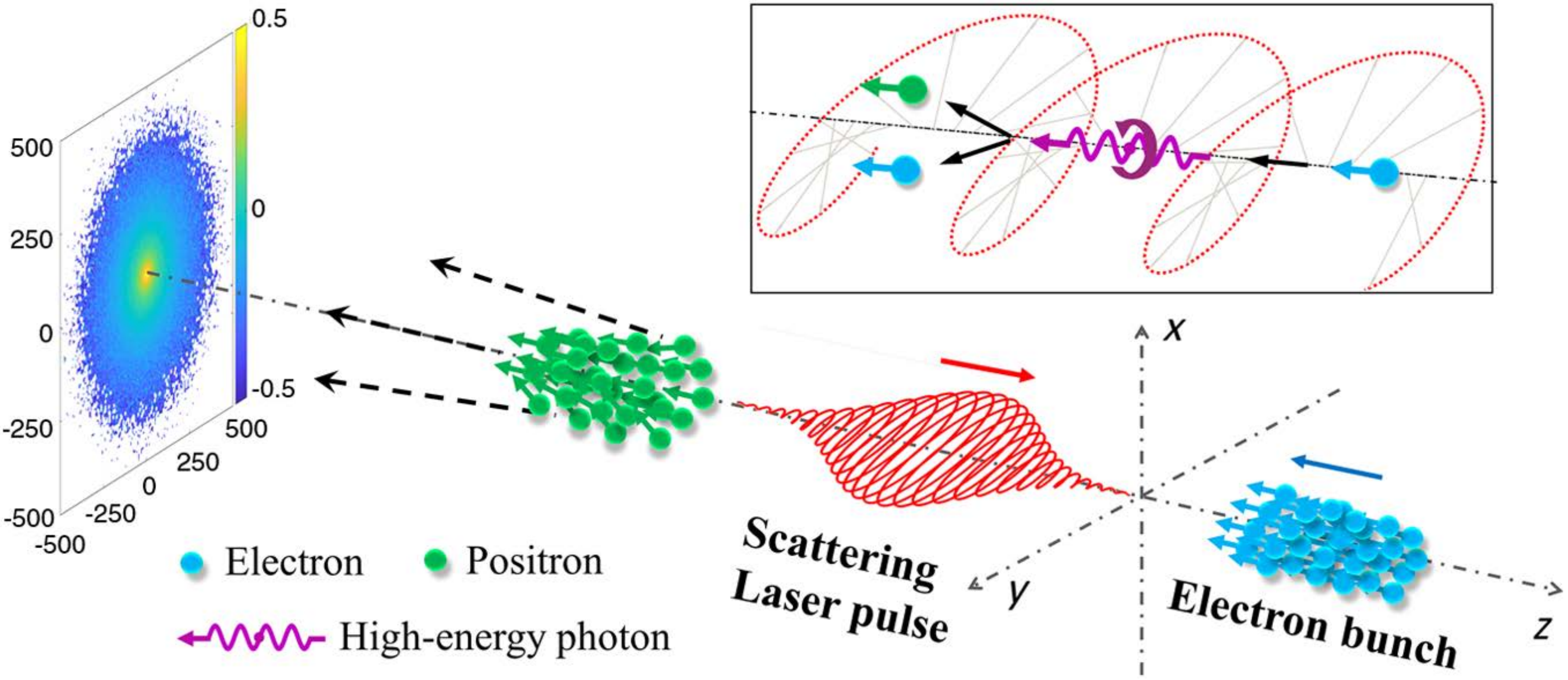}
	\caption{Scenarios of generation of a LSP ultrarelativistic positron beam via an ultraintense laser pulse head-on colliding with a counterpropagating LSP electron beam\cite{li2020production}.}\label{fig5-4}	
\end{figure}
The first scheme can be implemented by colliding an ultrarelativistic electron beam with a counter-propagating dichromatic petawatt laser pulse (see the upper sketch in Fig.~\ref{fig5-3}). First,  the pair-production probability with the positron spin parallel to the magnetic field is around 80$\%$, see Fig.~\ref{fig5-3}(b), so the direction of the spin of the positron is much more likely to be along that of the magnetic field during the pair-production process. Second, the probability of pair-production depends strongly on the laser fields, so the process takes place mainly in the dominant half-cycles (the magnetic field points in the $+y$ direction) of the two-color field and is fully suppressed in the weak half-cycle, see Fig.~\ref{fig5-3}(a). The latter is the main source of highly polarized positrons in a two-color field. According to the published literature \cite{chen2019polarized}, highly polarized positron beams (60\% polarization degree) can be produced at the femtosecond time-scale, with a small angular divergence ($\sim$74 mrad) and high density ($\sim$ 10$^{14}$ cm$^{-3}$). However, the effect of photon polarization is not resolved in this work. Subsequent studies  further explained the effect of photon polarization on the polarization and yield of the positron beam \cite{dai2022photon,wan2020high}.

Note that the produced positrons in the above proposals are transversely polarized. LSP positrons can be produced \cite{li2020production} in the second scheme, which can be implemented by colliding an ultraintense CP laser beam with a counter-propagating LSP electron beam, see Fig.~\ref{fig5-4}. In this scheme, CP $\gamma$-photons are produced via strongly NLC scattering, as LSP electrons propagate through the CP laser field \cite{li2020polarized}. Because the polarization transfer efficiency can reach nearly 100\% for the energetic positrons scattered into smaller angles, this scheme provides the feasible post-selection procedure to produce high-quality positron beams. A highly polarized (40\%–65\%), intense ($10^5-10^6$/bunch), collimated (5–70 mrad) positron beam can be generated this way on a femtosecond time-scale.

\begin{figure}[!t]
	\centering
	\includegraphics[width=0.9\linewidth]{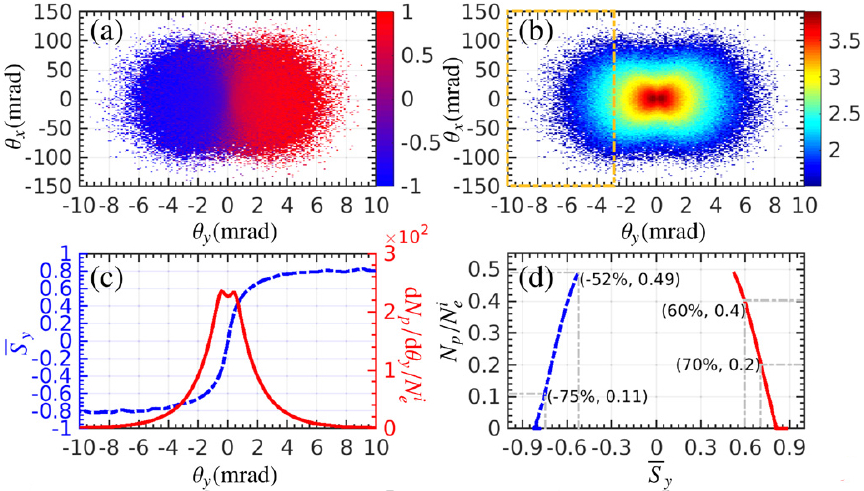}
	\caption{(a) and (b): Transverse distributions of the positron spin component $S_y$ and angular distribution of positron density($\mathrm{log}_{10}(\mathrm{d}^2N_p/\mathrm{d}\theta_x\mathrm{d}\theta_y/N_e^i)$). (c): Average spin $\overline{S_y}$ (blue-dashed curve) and positron density vs $\theta_y$. (d): $N_p/N_e^i$ vs $\overline{S_y}$. $N_p$ and $N_e^i$ are the number of positrons and primary electrons, respectively. $\theta_x = \mathrm{arctan}(P_{+,x}/P_{+,z})$, $\theta_y = \mathrm{arctan}(P_{+,y}/P_{+,z})$ is the positron deflection angles, respectively.\cite{wan2020ultrarelativistic}
}\label{fig5-5}
\end{figure}
In addition to the above two schemes, highly polarized positron beams can also be produced by the spin-splitting mechanism, which is implemented in the interaction of an ultrarelativistic electron beam with an EP laser beam \cite{wan2020ultrarelativistic}. For a laser field with a small ellipticity ($E_x\gg E_y$), the mechanism by which the positrons acquire polarization is mainly based on two facts (as in Ref. \cite{li2019ultrarelativistic}). The first is asymmetry of the fully spin-resolved pair-generation probabilities. Here, spins of most of the positrons will point in the direction of the magnetic field at the production time, that is, spins of the positrons produced for $E_x \textgreater 0$ will nearly point in the $+y$ direction, and vice versa. The second is the fact that the final transverse positron momenta mainly depend on the directions of the electromagnetic field components at the production time. For example,  the final transverse momenta of the positrons produced for $E_y \textgreater 0$ will be positive. Due to the fact that $E_x$ and $E_y$ always have the same sign in the EP laser, positrons in different spin states will split into two opposite directions [see Fig.~\ref{fig5-5}(a) and (b)]. A polarized positron beam can be obtained by implement a selection over $\theta_y$ [see Fig.~\ref{fig5-5}(c) and (d)], and the degree of polarization and yield of positrons are decreased by $\sim$35\% and $\sim$13\%, respectively, when the polarization of $\gamma$ photons is resolved \cite{dai2022photon}.

Another example of the spin-splitting scheme, which demonstrates feasibility of production of a densely-polarized positron beam, is implemented in the interaction of an ultraintense LP laser with a foil target \cite{song2021dense}. When the laser irradiates the target, a large number of energetic positrons will be produced in the skin layer of the solid-density plasma, immediately, and are then pushed into the deeper plasma and leave the laser field. Positrons produced in the positive and negative half-cycles will escape from the plasma in opposite directions.

\section{Summary and outlook}\label{sec:summ}

The theoretical investigations of the production of high-energy polarized particles have been reviewed in this paper. Acceleration of a polarized particle beam via laser-plasma interactions has the prominent advantages of compactness and high brilliance (see Sect. \ref{sec-polacc}). Therefore, the mechanism can significantly facilitate studies in materials science, nuclear physics and high-energy physics, while at the same time it opens the way for producing high-quality polarized $\gamma$-photons and positron beams via the strong-field QED processes. The spin and polarization are intrinsic particle properties which play important roles in the study of many QED processes, such as modification of the spectral structure and particle yield. Realization of the polarization build-up of the scattered particles, in the strong-field QED processes, requires dealing with peculiar configurations of beam and laser fields. Comprehensive studies have shown that high polarization can be built up in the strongly nonlinear processes by ultraintense laser fields.
In nonlinear Compton scattering, for example, the polarized electrons can be produced by an asymmetrically bichromatic laser field because of the spin-dependent photon radiation rates, and also by elliptically polarized laser fields due to the spin-dependent radiation-reaction (see Subsect. \ref{sec2-polele}). Unlike the radiative spin polarization of electrons, polarization of the $\gamma$-photons in nonlinear Compton scattering is transferred directly from either moderate-intensity laser photons or from electron spin, iprovided the laser field is ultraintense (see Sect. \ref{sec-polgamm}). In the nonlinear Breit-Wheeler process, driven by an ultraintense laser field, the positron polarization can be built up through either an asymmetrical laser field or transferred from the photon helicity (see Subsect. \ref{sec2-polpos}). Strong-field QED cascade processes driven by ultraintense lasers allow for the production of a QED plasma, which plays an essential role in high-energy studies in astro-particle physics. The related polarization effects are also briefly summarized (see Subsect. \ref{sec2-polgamm}).

The strong-field QED processes can be investigated in future experiments employing currently available and future petawatt (and possibly beyond) laser facilities. For example, the constructing 10-petawatt laser beams in ELI-NP facility will enable the nonlinear QED experiments by the production of multi-GeV electrons and multi-100 MeV protons \cite{turcu2016high,ataman2017experiments,blackburn2020radiation}.
Taking advantage of the produced high-energy highly-polarized particle sources in the regime of strong-field QED, for example, it should be possible to design and build polarized photon-photon colliders to experimentally realize processes like the vacuum birefringence, two-photon BW pair-production and photon-photon elastic scattering, such that the characteristic polarization-associated signatures may be distinguished from the background noise. Furthermore, with the ultraintense laser-driven QED plasmas being of strong relevance in the investigation of high-energy astrophysics \cite{zhang2020relativistic}, the understanding of QED cascades, especially the polarization effects involved, in the production of the laser-driven QED plasma is a requisite for future experimental studies.

\bmhead{Acknowledgments}
This work is supported by the National Natural Science Foundation of China (Grants Nos. 12022506, 11874295, 12105217, 11905169, 11875307 and 11935008). The work of YIS is supported by an American University of Sharjah Faculty Research Grant (FRG21).
\bibliography{ref-polpart}% common bib file


\end{document}